\documentclass[12pt]{article}\usepackage[]{graphicx}\usepackage[]{xcolor}
\makeatletter
\def\maxwidth{ %
  \ifdim\Gin@nat@width>\linewidth
    \linewidth
  \else
    \Gin@nat@width
  \fi
}
\makeatother

\definecolor{fgcolor}{rgb}{0.345, 0.345, 0.345}
\newcommand{\hlnum}[1]{\textcolor[rgb]{0.686,0.059,0.569}{#1}}%
\newcommand{\hlcom}[1]{\textcolor[rgb]{0.678,0.584,0.686}{\textit{#1}}}%
\newcommand{\hlopt}[1]{\textcolor[rgb]{0,0,0}{#1}}%
\newcommand{\hlstd}[1]{\textcolor[rgb]{0.345,0.345,0.345}{#1}}%
\newcommand{\hlkwb}[1]{\textcolor[rgb]{0.69,0.353,0.396}{#1}}%
\newcommand{\hlkwc}[1]{\textcolor[rgb]{0.333,0.667,0.333}{#1}}%
\newcommand{\hlkwd}[1]{\textcolor[rgb]{0.737,0.353,0.396}{\textbf{#1}}}%

\usepackage{framed}
\makeatletter
\newenvironment{kframe}{%
 \def\at@end@of@kframe{}%
 \ifinner\ifhmode%
  \def\at@end@of@kframe{\end{minipage}}%
  \begin{minipage}{\columnwidth}%
 \fi\fi%
 \def\FrameCommand##1{\hskip\@totalleftmargin \hskip-\fboxsep
 \colorbox{shadecolor}{##1}\hskip-\fboxsep
     \hskip-\linewidth \hskip-\@totalleftmargin \hskip\columnwidth}%
 \MakeFramed {\advance\hsize-\width
   \@totalleftmargin\z@ \linewidth\hsize
   \@setminipage}}%
 {\par\unskip\endMakeFramed%
 \at@end@of@kframe}
\makeatother

\definecolor{shadecolor}{rgb}{.97, .97, .97}
\definecolor{messagecolor}{rgb}{0, 0, 0}
\definecolor{warningcolor}{rgb}{1, 0, 1}
\definecolor{errorcolor}{rgb}{1, 0, 0}
\newenvironment{knitrout}{}{} 

\usepackage{alltt}

\usepackage{graphicx}
\usepackage{amsmath,amssymb} 
\usepackage{tikz}
\usepackage[linesnumbered]{algorithm2e}
\usepackage{setspace}
\usepackage{multirow}

\usepackage{fancyvrb} 
\usepackage{verbatim}

\def\cocoa{{\hbox{\rm C\kern-.13em o\kern-.07em C\kern-.13em o\kern-.15em A}}}
\DeclareMathOperator{\sign}{sign}

\newcounter{ejemplo}

\newtheorem{theorem}{Theorem}

\newtheorem{example}[ejemplo]{Example}

\newtheorem{lemma}[theorem]{Lemma}

\newenvironment{proof}[1][Proof]{\textbf{#1.} }{\ \rule{0.5em}{0.5em}}

\providecommand{\keywords}[1]
{
  \small	
  \textbf{\textit{Keywords---}} #1
}

\title{Lasso and elastic nets by orthants}
\author{H. Maruri-Aguilar}
\IfFileExists{upquote.sty}{\usepackage{upquote}}{}
\begin{document}
\maketitle

\begin{abstract}
We propose a new method for computing the lasso path,
using the fact that 
the Manhattan norm of the coefficient vector is linear
over every orthant of the parameter space.
We use simple calculus and present an algorithm  
in which the lasso path is series of orthant moves. Our proposal gives the same
results as standard literature, with the advantage of neat 
interpretation of results and explicit lasso formul\ae. We 
extend this proposal to elastic nets and obtain explicit, 
 exact formul\ae{} for the elastic net path, and with a simple change, our lasso algorithm can be used 
for elastic nets.
We present computational examples and provide simple 
\texttt{R} prototype code.
\end{abstract}


\tableofcontents

\section{Introduction}\label{sec_intro}

This paper is concerned with penalised estimation for 
the linear regression model 
\begin{equation}\label{ec_regression}\mathbf{Y}=\mathbf{X}\beta+\epsilon.\end{equation}
The vector of response values is $\mathbf{Y}$, 
the parameter vector is $\beta$; the covariate 
matrix  $\mathbf{X}$ has
$p$ linearly independent columns and $n$ rows, one for every observation
and $\epsilon$ is the vector of normal independent error terms with zero mean 
and variance $\sigma^2$.
In this paper, $\mathbf{Y}$ and $\mathbf{X}$ refer to the observed
response vector and covariate matrix, respectively.
Following lasso practice, both $\mathbf{Y}$ and columns of $\mathbf{X}$ are 
centered around their sample means so the regression does not have intercept term.

\subsection{Lasso regularization}
For parameter estimation of model (\ref{ec_regression}), 
the Lasso \cite{T1996}
minimizes the criterion 
\begin{equation}\label{eclasso}
L=\frac{1}{2}||\mathbf{Y}-\mathbf{X}\beta||_2^2+\lambda||\beta||_1,
\end{equation}
where $||\cdot||_2$ and $||\cdot||_1$ are the Euclidean and
Manhattan norms.
For fixed 
$\lambda\geq 0$, the lasso estimate $\hat\beta=\hat\beta(\lambda)$ 
minimizes $L$ over 
$\mathbb{R}^p$.
 As $\lambda$ 
increases, $\hat\beta(\lambda)$ shrinks  
from the least squares $\hat\beta(0)=(\mathbf{X}^T\mathbf{X})^{-1}\mathbf{X}^T\mathbf{Y}$
towards 
zero. 
By this shrinking feature, lasso works as a continuous 
method for subset selection \cite{HTF2009}.  

The criterion $L$ of Equation (\ref{eclasso}) is convex,
but its second term makes the estimates nonlinear as 
function of the observations, and apart from the case where $\mathbf{X}$ has
orthogonal columns, there is no general closed formula for the 
estimate $\hat\beta(\lambda)$, see \cite{T1996,HTF2009}.
Lasso estimation is a quadratic programming problem,
and the solution has been computed with a variation of 
Least Angle Regression (LAR) \cite{EHJT2004,TT2011}. 
Further developments on lasso estimation are the use of a descent algorithm
and homotopy as well as
 the use of duality \cite{OPT2000,OPT2000b}. 
Lasso has been studied with Bayesian principles
 for experimental screening using Laplace priors
for the parameters \cite{NOY2015} and using prior information for
generalized linear models \cite{JHZ2016}. Our paper does not
use Bayesian priors and is based on simple ideas that we describe next.

\subsection{Contributions of this paper}\label{sec_contrib}

We solve lasso estimation by noting that over
 every orthant, the Manhattan norm
 $||\beta||_1$ is a linear function of $\beta$ . This allows the use of standard 
calculus
to maximize $L$, considering the part of $L$ in
a given orthant as defined in Equation (\ref{eclassoort}).
By construction, our proposal gives the exact minimization of $L$ and no approximations 
are performed. 
We obtain exact, explicit formul\ae{} for $\hat\beta$ that minimizes the lasso
criterion and propose an algorithm to compute the lasso path.

We also analyse elastic nets using orthants. The  
 criterion to be minimized is 
\begin{equation}\label{ecnet}E=\frac{1}{2}||\mathbf{Y}-\mathbf{X}\beta||_2^2+\lambda\alpha||\beta||_1+\lambda\frac{1-\alpha}{2}||\beta||_2^2.
\end{equation}
For our analyses, we define the part of $E$ in a given orthant in Equation (\ref{eenet}).
Similar to our lasso development, we minimize $E$ and 
obtain explicit formul\ae{} for $\hat\beta$. A simple change to 
our lasso algorithm allows the computation for elastic nets.

The gradients of criteria $L$ and $E$ are known in the statistical
literature, see \cite{FHHT2007} and \cite{FTH2010}. Those computations are performed 
at non-zero estimates of the net path, and coefficient
updates are based upon computation of partial residuals and use a soft thresholding operator. 
Our method is 
different and simpler; criteria $L$ and $E$ are split in orthants using the local versions
which are quadratic forms and only require elementary computations of
multivariate calculus. 

\subsection{Order of the paper}

We introduce the orthant method to lasso in Section \ref{sec_method}.
We start with a single parameter and then 
develop the
 multiple parameter case using orthants in Section \ref{sec_multiparam} and 
apply standard calculus to obtain 
 the minimizer $\hat\beta(\lambda)$.
 Although the lasso method is a particular instance of the elastic net 
of Section \ref{sec_en}, we present lasso first as it is a
simpler case with linear trajectories as function of $\lambda$ and it is 
also more stable for computations, when compared 
with elastic net. 

In Section \ref{sec_path} construct the lasso
path, which is the collection of $\hat\beta(\lambda)$
that minimize $L$. The $\hat\beta(\lambda)$ are piecewise linear functions of $\lambda$, 
that change orthant at certain values of $\lambda$ known as breakpoints. 
We discuss an all orthant approach in Section \ref{sec_orthant}
and present our main algorithm in Sections \ref{sec_adaptive}
and \ref{sec_theactualalgorithm}.
The algorithm obtains breakpoints of the 
path, which are exit and entry points when moving between orthants. 
We give a detailed example of the algorithm in Section \ref{sec_example}.

In Section \ref{sec_en} we study elastic net with orthants. This development
mirrors what we did in lasso, with the important difference that the coefficient
trajectories are nonlinear, piecewise functions of $\lambda$. Despite this, the 
computation of 
the elastic net path still consists in determining exit and entry points for
orthants and thus for elastic nets we use the lasso algorithm, with a simple change
that we describe.

A discussion of results is presented in Section \ref{sec_final}. We
comment upon the \texttt{lars} implementation \cite{HE2022} and
our algorithm in Section \ref{sec_lassocomp}.
In Section \ref{sec_solution} we discuss the numerical 
solution at the core of net orthant method, and in 
Section \ref{sec_implementation} we compare
the results 
between our proposal and the \texttt{glmnet} implementation \cite{FTH2010}.
This is done both by examples and with a simulation study. 
Finally, in Section \ref{sec_furtherwork} we survey future work directions to our
method. This paper has an Appendix with proofs, examples and \texttt{R} code
prototype for lasso and elastic nets. 

\section{Lasso by orthants}\label{sec_method}

The Lasso criterion $L$ of Equation (\ref{eclasso}) is
a convex function which, apart from  $\lambda=0$, cannot be written as 
a quadratic form over the full range of
potential parameter values $\mathbb R^p$.
However, if we consider $L$ over orthants
in $\mathbb R^p$, in each orthant the problem
is a quadratic form for which simple closed formul\ae{} are available.
We start with 
one parameter and then describe the general methodology.

\subsection{Single parameter case}\label{sec_single}

This section is similar to part of the one parameter 
development of Equation (3) and following text in \cite{FHHT2007}. However our 
treatment is simpler and we do not require standardized variables nor
use concepts like soft thresholding.

Consider $\mathbf{X}$ with a single column,
i.e. $\mathbf{X}=(x_{11}\;x_{21}\cdots\;x_{n1})^T$ with 
model parameter $\beta_1$.
The Lasso criterion is
$L=\frac{1}{2}\sum_{i=1}^n\left(y_i-x_{i1}\beta_1\right)^2+\lambda|\beta_1|$, to be minimized 
for
$\beta_1$ over $\mathbb R$.
Trivially, for $\beta_1>0$, the absolute value $|\beta_1|$ equals $\beta_1$, while for
$\beta_1<0$, we have $|\beta_1|=-\beta_1$. Each of these cases is one orthant of the
real line $\beta_1\in\mathbb R$. To complete the full range of $\beta_1$ we 
add the lower dimensional orthant $\beta_1=0$ thus decomposing
$\mathbb{R}=(-\infty,0)\bigcup\{0\}\bigcup(0,\infty)$ and rewriting $L$ as
\[
L=\left\{\begin{array}{lr}
\frac{1}{2}\sum_{i=1}^n\left(y_i-x_{i1}\beta_1\right)^2+\lambda\beta_1&\mbox{ if }\beta_1>0\\
\frac{1}{2}\sum_{i=1}^ny_i^2&\mbox{ if }\beta_1=0\\
\frac{1}{2}\sum_{i=1}^n\left(y_i-x_{i1}\beta_1\right)^2-\lambda\beta_1&\mbox{ if }\beta_1<0\\
\end{array}\right.
\]

The formulation above 
turns the minimization of $L$
in a simple quadratic problem with a closed form solution for every orthant.
The path is a collection of orthant moves starting at $\lambda=0$ 
with  least squares $\hat\beta_1(0)$, whose sign
determines the initial orthant. 
The lasso path proceeds in the direction of steepest descent
and we find $\lambda$ at which the trajectory moves
to a neighboring orthant.

\begin{table}\begin{center}
\begin{tabular}{cc}
\begin{tabular}{rrrr}
$\mathbf{X}_1$&$\mathbf{X}_2$&$\mathbf{X}_3$&$\mathbf{Y}$\\\hline
0&0&-1&1\\
-1&1&0&1\\
0&-1&-1&0\\
-1&0&0&-1\\
-1&1&0&1\\
-1&-1&-1&1\\
4&0&3&-3\\
\end{tabular}\;\;&\;\;
\begin{tabular}{rrrr}
$\mathbf{X}_1$&$\mathbf{X}_2$&$\mathbf{X}_3$&$\mathbf{Y}$\\\hline
-1&1&0&1\\
-1&1&-1&1\\
0&0&-1&0\\
0&1&-1&-1\\
1&-1&1&0\\
1&-2&2&-1\\
\end{tabular}\\(a)&(b)\end{tabular}\end{center}
\caption{Simulated data for (a) Example \ref{ex_1d3} and (b) Example \ref{exlarso}.}\label{tab_ex1d}
\end{table}

For $\lambda=0$ the estimate  is
 $\hat \beta_1(0)=\sum_{i=1}^nx_{i1}y_i/\sum_{i=1}^nx_{i1}^2$ 
with  orthant depending on the sign of
$\sum_{i=1}^nx_{i1}y_i$. If $\sum_{i=1}^nx_{i1}y_i>0$, 
the path proceeds over orthant $\beta_1>0$ as 
 $\hat \beta_1(\lambda)=\left(\sum_{i=1}^nx_{i1}y_i-\lambda\right)/\sum_{i=1}^nx_{i1}^2$
 that minimizes $L$ 
for  $\lambda>0$. 
When $\lambda=\sum_{i=1}^nx_{i1}y_i$, the estimate becomes
$\hat\beta_1=0$ at which point the solution leaves the orthant
$\beta_1>0$. As the estimate has shrunk to zero, the
lasso path ends. If 
$\sum_{i=1}^nx_{i1}y_i<0$, 
then $\hat \beta_1(0)<0$ and the 
path is $\hat \beta_1(\lambda)=\left(\sum_{i=1}^nx_{i1}y_i+\lambda\right)/\sum_{i=1}^nx_{i1}^2$ 
which shrinks to zero when $\lambda=\left|\sum_{i=1}^nx_{i1}y_i\right|$.
We give an example.

\begin{example}\label{ex_1d3}
Using columns $\mathbf{Y}$
and $\mathbf{X}_1$ of Table \ref{tab_ex1d}(a)
as response and explanatory variable,
a regression
model with parameter $\beta_1$ is considered.
For $\lambda=0$ we have $\hat\beta_1(0)=-0.7$,
located in orthant $\beta_1<0$, where the path starts. This is
because $\sum_{i=1}^nx_{i1}y_i=-14<0$, and the Lasso path is
$\hat\beta_1(\lambda)=(-14+\lambda)/20$.
For increasing values of $\lambda$, $\hat\beta_1(\lambda)$ shrinks towards zero and 
when $\lambda\geq \left|\sum_{i=1}^nx_{i1}y_i\right|=14$, the estimate is 
zero. 
Figure \ref{figex1d} (Appendix) shows the criterion $L$ for this example and four
values of $\lambda$. The location of the lasso estimate $\hat\beta_1$ is indicated 
by dashed lines.
\end{example}

\subsection{Multiple parameters}\label{sec_multiparam}

In the unidimensional case, the parameter region 
was split into three orthants as $\mathbb{R}=(-\infty,0)\bigcup\{0\}\bigcup(0,\infty)$.
We extend this idea with the 
Kronecker product of unidimensional orthants 
$\mathbb{R}^p=\bigotimes_{i=1}^p\mathbb{R}=\bigotimes_{i=1}^p\left((-\infty,0)\bigcup\{0\}\bigcup(0,\infty)\right).$
Each of the $3^p$ disjoint orthants is the interior of a polyhedral cone, which
is identified with a vector of $p$ numbers taken
from $\{\pm1,0\}$. This vector compose the entries of a diagonal matrix of size $p$, that we refer to as $\mathbf{C}$. Using $\mathbf{C}$, vectors inside
an orthant are $\mathbf{C}\mathbf{u}$, where $\mathbf{u}$ is a positive vector, that 
is $\mathbf{u}\in \mathbb R^p_{>0}$. The matrix $\mathbf{C}$ is central in this work, and
we refer to the orthant determined by $\mathbf{C}$ as ``orthant $\mathbf{C}$''.

As example, consider $\beta_1>0,\beta_2=0,\beta_3<0,\beta_4>0$.
This orthant is 
$(0,\infty)\bigotimes\{0\}\bigotimes(-\infty,0)\bigotimes(0,\infty) =
 \{\mathbf{C}\mathbf{u}:\mathbf{u}\in \mathbb R^4_{>0}\}$ with $\mathbf{C}=\mbox{diag}(1,0,-1,1)$. 
 
We refer to orthants with symbols \texttt{+}, \texttt{-}, \texttt{0}
for the diagonal of $\mathbf{C}$ so that e.g. \texttt{+0-+} refers to the orthant 
with $\mathbf{C}=\mbox{diag}(1,0,-1,1)$.
We consider strict inequalities for non-zero elements so that the orthants are disjoint. If 
the analysis requires non-strict inequalities, this is achieved by considering all disjoint orthants
involved. For example, if the desired region was $\beta_1\geq 0,\beta_2\leq 0,\beta_3= 0$, we would 
consider the orthants \texttt{000}, \texttt{+00},
\texttt{0-0} and \texttt{+-0}.

We formulate the parameter vector and lasso criterion 
 over an orthant determined by matrix $\mathbf{C}$. In orthant $\mathbf{C}$, the 
 parameter vector $\beta$ is  
\begin{equation}\beta=\mathbf{C}\mathbf{u},\label{ecbeta}\end{equation}
with $\mathbf{u}\in \mathbb R^p_{>0}$, 
and 
the Lasso criterion of Equation (\ref{eclasso}) becomes
\begin{equation}\label{eclassoort}L_\mathbf{C}=\frac{1}{2}\mathbf{Y}^T\mathbf{Y}
-\mathbf{u}^T \mathbf{C}\mathbf{X}^T\mathbf{Y}
+\frac{1}{2}\mathbf{u}^T \mathbf{C}\mathbf{X}^T\mathbf{X}\mathbf{C}\mathbf{u}
+\lambda \mathbf{u}^T \mathbf{C}^2 \mathbf{1},\end{equation}
where the symbol $\mathbf{1}$ is the vector of ones of dimension $p\times 1$.

We have just turned the Lasso criterion (\ref{eclasso}) into a standard quadratic form (\ref{eclassoort}) by
considering orthants. 
The lasso penalization $||\beta||_1=\sum_{i=1}^p|\beta_i|$ of 
Equation (\ref{eclasso}) becomes $\beta^T\mathbf{C}\mathbf{1}=\mathbf{u}^T \mathbf{C}^2 \mathbf{1}$ in (\ref{eclassoort}). 
 The  
quantity $\mathbf{u}^T \mathbf{C}^2 \mathbf{1}$ is non-negative, as it is the sum of positive elements in $\mathbf{u}$, and $\mathbf{C}^2\mathbf{1}$ automatically 
considers zeroes as needed by the current orthant through its matrix $\mathbf{C}$. For example
with
\texttt{+0-+0-0} we have $\mathbf{u}^T \mathbf{C}^2 \mathbf{1}=u_1+u_3+u_4+u_6>0$ because the $u_i$ are all positive.

In orthant $\mathbf{C}$, the quadratic form $L_\mathbf{C}$ is well formulated because of linear independence of columns in $\mathbf{X}$.
When we merge orthants, we recover $L$ over all of $\mathbb R^p$
and our development keeps the continuity and convexity properties of $L$.

\subsection{Estimation of $\beta$ per orthant}

We next obtain closed form solutions for the minimization of 
Equation (\ref{eclassoort}) using standard calculus techniques. 
Our development gives a simple interpretation to the lasso estimates. With 
 careful handling of the minimization solutions, we reconstruct the lasso path
 in Sections \ref{sec_orthant} and \ref{sec_adaptive}.

To minimize $L_\mathbf{C}$, the vector of derivatives
with respect to entries in $u$ is
\[\frac{\partial L_\mathbf{C}}{\partial \mathbf{u}} = - \mathbf{C}\mathbf{X}^T\mathbf{Y}
+ \mathbf{C}\mathbf{X}^T\mathbf{X}\mathbf{C}\mathbf{u}
+ \lambda \mathbf{C}^2 \mathbf{1}.\]
Although $L_\mathbf{C}$ is to be evaluated only with positive $\mathbf{u}$, note that $L_\mathbf{C}$ is a quadratic
form not formally constrained to this domain, and its derivative is well defined. 
To determine critical points, the gradient is set to zero so that over the orthant 
determined by $\mathbf{C}$, the vector $\hat{\mathbf{u}}$
that minimizes $L_\mathbf{C}$ satisfies the linear system 
\begin{equation}\label{ec_linearsystem}
\mathbf{C}\mathbf{X}^T\mathbf{X}\mathbf{C}\hat{\mathbf{u}} =\mathbf{C}\mathbf{X}^T\mathbf{Y}
- \lambda \mathbf{C}^2 \mathbf{1}.\end{equation}
When $\mathbf{C}$ is a full 
rank matrix, this is a standard linear system in $\mathbf{u}$. Depending on the number of 
of non-zero entries in the diagonal of $\mathbf{C}$, the system may have less than $p$ 
active equations, with the non-active equations becoming tautologies of the type $0=0$ 
with no influence on the analysis. 

In what follows, we show
that the active equations form a solvable, square linear system.
We require a Lemma about the orthant matrix $\mathbf{C}$ and a
theorem stating a property of the generalized inverse 
of $\mathbf{C}\mathbf{X}^T\mathbf{X}\mathbf{C}$.
The proof of the Lemma is direct and it is not given, while
the proof of the Theorem is in Appendix 1.

\begin{lemma}\label{lemmaC}
Let $\mathbf{C}$ be a square diagonal matrix with 
entries from $\pm 1,0$. Then $\mathbf{C}$ equals its generalized
inverse $\mathbf{C}^-$. Furthermore, $\mathbf{C}^3=\mathbf{C}$.
\end{lemma}

\begin{theorem}\label{sinverse}
Let $\mathbf{S}:=\mathbf{C}\mathbf{X}^T\mathbf{X}\mathbf{C}$, 
where $\mathbf{C}$ is a diagonal matrix with entries from $0,\pm 1$. 
The generalized inverse $\mathbf{S}^-$ of $\mathbf{S}$ satisfies
$\mathbf{S}\mathbf{S}^-=\mathbf{S}^-\mathbf{S}=\mathbf{C}^2$.
\end{theorem}

Using Theorem \ref{sinverse}, the solution of the system (\ref{ec_linearsystem}) is
\begin{equation}\label{sols}\mathbf{C}^2\hat{\mathbf{u}} =\mathbf{S}^-\left(\mathbf{C}\mathbf{X}^T\mathbf{Y}- \lambda\mathbf{C}^2  \mathbf{1}\right).\end{equation}
Equation (\ref{sols}) is
the equation of a line with starting point $\mathbf{S}^-\mathbf{C}\mathbf{X}^T\mathbf{Y}$ and direction
$-\lambda{}\mathbf{S}^-\mathbf{C}^2 \mathbf{1}$, which 
points in the direction of maximum change of the solution. To be considered as
a valid trajectory, $\mathbf{C}^2\hat{\mathbf{u}}$ should be positive for all its components. 
Not every $\mathbf{C}$ gives a valid solution, and we later
describe how to determine which orthants $\mathbf{C}$ correspond to the 
 solution of the lasso minimization.

Although the vector $\mathbf{C}^2\hat{\mathbf{u}}$ has dimension $p\times 1$, the
non-trivial entries on it correspond to non-zero elements in the diagonal of $\mathbf{C}$,
that is non zero entries in $\mathbf{C}^2\hat{\mathbf{u}}$ are those for which the diagonal entry in $\mathbf{C}$ is
one of $\pm 1$. 
That is, the matrix $\mathbf{C}^2$ in Theorem \ref{sinverse} and following developments is 
like an identity matrix that, depending on the entries of $\mathbf{C}$, 
may contain some zero elements in its diagonal.  
 An example of computation of $\mathbf{S}^-$ is given in Appendix 2.

The lasso estimate is 
obtained by left multiplying Equation (\ref{sols}) by $\mathbf{C}$
and using $\mathbf{C}^3=\mathbf{C}$ of Lemma \ref{lemmaC}, that is 
$\hat\beta=\mathbf{C}\mathbf{C}^2\hat{\mathbf{u}}=\mathbf{C}\hat{\mathbf{u}}$.
The estimate is 
\begin{equation}\hat \beta=\mathbf{C}\mathbf{S}^-\mathbf{C}\left(\mathbf{X}^T\mathbf{Y}- \lambda \mathbf{C}\mathbf{1}\right),\label{ec_beta}\end{equation}
which is composed of a linear function of 
observations $\mathbf{C}\mathbf{S}^-\mathbf{C}\mathbf{X}^T\mathbf{Y}$ and a biasing term 
$-\lambda{}\mathbf{C}\mathbf{S}^-\mathbf{C}^2\mathbf{1}$
that does not depend on observations. Because
of this, the lasso estimate $\hat\beta$ is a nonlinear function of $\mathbf{Y}$. 
With orthonormal $\mathbf{X}$, $\hat\beta$ of Equation (\ref{ec_beta}) equals the 
lasso estimator of Equation (3) in \cite{T1996}.
We use Equation (\ref{ec_beta}) to
evaluate $\hat\beta$ in a segment of the lasso path, which we show next.

\begin{example}\label{ex_1d3bis} The initial lasso path of Example \ref{ex_1d3} is retrieved 
with Equation (\ref{ec_beta}) by 
noting that $\mathbf{X}^T\mathbf{X}=20$ so 
$\mathbf{S}=\mathbf{C}\mathbf{X}^T\mathbf{X}\mathbf{C}=20$ and $\mathbf{S}^-=1/20$ because we are in orthant 
\texttt{-} and $\mathbf{C}=-1$. We 
use $\mathbf{X}^T\mathbf{Y}=-14$ to write $\hat\beta=-1/20\cdot(14-\lambda)$.
\end{example}

\begin{example}\label{exlarso}
For the data of Table \ref{tab_ex1d}(b) and $\lambda=0$, the least squares estimate 
of $\beta$ is $(-1.25, -0.3333, 0.0833)^T$
which is the first term of Equation (\ref{ec_beta}) and is the starting point of the 
lasso trajectory in the orthant  \texttt{--+}. 
In this initial orthant, 
the lasso path is 
$\hat\beta=(-1.25, -0.3333, 0.0833)^T-\lambda\cdot (-2, -1.6667, -0.3333)$. 
\end{example}

Apart from the beginning of the path, the
first term of (\ref{ec_beta}) may not lie inside the orthant $\mathbf{C}$, and only when adding the 
second term,
$\hat\beta$ lies in $\mathbf{C}$. This 
has to be checked, that is, 
for given $\mathbf{C},\lambda$, the coefficient $\hat\beta=\mathbf{C}\hat{\mathbf{u}}$ of Equation (\ref{ec_beta}) 
will only be in its $\mathbf{C}$-orthant when all the components of 
$\mathbf{C}^2\hat{\mathbf{u}}$ of
Equation (\ref{sols}) are non-negative. 
This is a consequence of our development, where we computed  
for positive $\mathbf{u}$ and moved back to the corresponding orthant by left 
multiplying by $\mathbf{C}$. 
The following example uses Equation (\ref{ec_beta}) at an intermediate orthant 
in the path.

\begin{figure}[h]
\begin{center}
\begin{knitrout}
\definecolor{shadecolor}{rgb}{0.969, 0.969, 0.969}\color{fgcolor}
\includegraphics[width=\maxwidth]{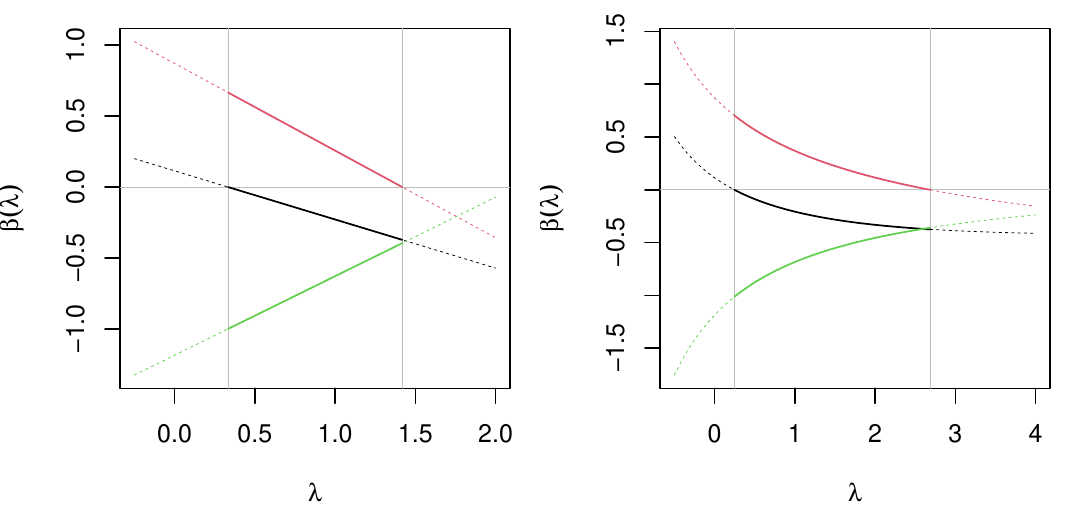} 
\end{knitrout}
\end{center}
\caption{(left) Lasso trajectories of Example \ref{ex_lapath};  
(right)
 elastic net trajectories
of Example \ref{ex_netpath}. The plots also 
show that Equations (\ref{ec_beta}) and (\ref{ecsolsnet}) can
 be evaluated for $\lambda<0$.
Line colors black, red and green are for $\beta_1,\beta_2,\beta_3$,
respectively; grey lines indicate example values for $\lambda$ and the 
zero line.}\label{fig_lapath}
\end{figure}

\begin{example}\label{ex_lapath}
For the data of Table \ref{tab_ex1d}(a),
when 
$0.333<\lambda<1.419$, 
the lasso path 
crosses through orthant \texttt{-+-}. 
The trajectory is computed with Equation (\ref{ec_beta}) yielding 
$\hat\beta=(0.1143, 0.8714, -1.1857)^T-\lambda\cdot(0.3429, 0.6143, -0.5571)^T$.
None of two terms are in \texttt{-+-}, but the sum 
lies in this orthant over the range of $\lambda$. 
The trajectory is
plotted in Figure \ref{fig_lapath}, where solid lines show the transit of $\hat\beta$
through \texttt{-+-}. Outside the range of $\lambda$, 
the trajectories can still be computed,  
although these are not part of lasso path and are indicated with dotted lines
in the figure.
\end{example}

In summary, Equation (\ref{ec_beta}) is the explicit formula for the lasso path, but it 
has to be linked with a suitable orthant $\mathbf{C}$ and range for $\lambda$. 
The following section discusses the computation of the
path relative to orthant $\mathbf{C}$ and values of $\lambda$.

\section{Lasso trajectory by orthants}\label{sec_path}

For a matrix $\mathbf{C}$ and a value $\lambda$, 
substitution of $\hat\beta=\mathbf{C}\hat{\mathbf{u}}$ into the criterion $L_\mathbf{C}$ of 
Equation (\ref{eclassoort}) gives the smallest value of $L_\mathbf{C}$. This 
value is a polynomial function of degree two in $\lambda$ 
and we refer to it as $\hat L_\mathbf{C}$:
\begin{equation}
\hat L_\mathbf{C}=\frac{1}{2}\left( \mathbf{Y}^T\mathbf{Y}
-\mathbf{Y}^T\mathbf{X}\mathbf{C}\mathbf{S}^-\mathbf{C}\mathbf{X}^T\mathbf{Y} 
+2\lambda \mathbf{Y}^T\mathbf{X}\mathbf{C}\mathbf{S}^-\mathbf{1}
-\lambda^2 \mathbf{1}^T\mathbf{S}^-\mathbf{1}\right).
\label{ec_criterionopt}
\end{equation}
This formula can be evaluated for any pair $\mathbf{C},\lambda$, but not all 
evaluations of $\hat L_\mathbf{C}$ will correspond to
a path minimizing $L$. In what
follows, we reconstruct the lasso path by first presenting an exhaustive approach and then
the recommended algorithm.

\subsection{All orthants lasso analysis}\label{sec_orthant}

A simple approach to compute the estimate
$\hat\beta(\lambda)$ at a given $\lambda$ is to
evaluate $\hat L_\mathbf{C}$ over all orthants.
This exhaustive method considers all cases for the 
diagonal of $\mathbf{C}$ from $\bigotimes_{i=1}^p\{-1,0,1\}=\{-1,0,1\}^p$ and
 excludes those cases of $\mathbf{C}^2\hat{\mathbf{u}}$ that have 
one or more negative entries which implies that $\hat\beta$ is outside 
the orthant determined by $\mathbf{C}$.
After removing unfeasible cases, we select the orthant $\mathbf{C}$ over
which $\hat L_\mathbf{C}$ of Equation (\ref{ec_criterionopt}) is minimized
and retrieve the corresponding  lasso estimate $\hat\beta$.

The analysis for a single value of $\lambda$ turns directly into the coefficients in
the lasso path as follows. Set $\Lambda$ to be a collection
of $\lambda$ values of interest which are
 positive and no larger than
$\max\{ |(\mathbf{X}^T\mathbf{Y})_i|,i=1,\ldots,p\}$, i.e. the value at which all the
trajectories shrink to zero \cite{T1996}. 
In the earlier expression $(\cdot)_i$ means the $i-$th element of the argument.
For each $\lambda\in \Lambda$, select $\hat\beta$ associated
with $\mathbf{C}$ that minimizes
(\ref{ec_criterionopt}) over all orthants $\mathbf{C}$.
This collection of $\hat\beta$ is the lasso path for $\lambda\in \Lambda$. 

The exploration computes 
 $\hat\beta$ and $\hat L_\mathbf{C}$ for $3^p$ cases of orthants $\mathbf{C}$.
The advantage is that we retrieve the minimizer,
but a 
big drawback is its cost $3^p\# \Lambda$, and 
apart from small values of $p$, we would not advise to use it in general.

\subsection{Sequential lasso 1: Two types of moves}\label{sec_adaptive}

Assume that for a given $\lambda$, we are in orthant $\mathbf{C}$ 
and that by changing $\lambda$, we want to move to a different  
orthant in the path. 
Equation (\ref{sols}) suggests two possible moves available for us in the 
Lasso path: shrinkage and reactivation. 

\subsubsection{Shrinkage}\label{sec_shrinkage}

Starting from orthant $\mathbf{C}$, a list of candidate $\lambda$ values for shrinking
coefficients is given by those
values of $\lambda$ that make the coordinates of the trajectory 
of Equation (\ref{sols}) take value zero. At $i-$th coordinate,
this occurs for a value $\lambda^*_i$ computed as 
\begin{equation}\lambda^*_i=
\frac{(\mathbf{S}^-\mathbf{C}\mathbf{X}^T\mathbf{Y})_i}{(\mathbf{S}^- \mathbf{C}^2\mathbf{1})_i}.\label{ec_lambda}\end{equation}
This computation is done for $i$ in $1,\ldots,p$ such that 
the $i-$th diagonal element of the current orthant $\mathbf{C}$ is not zero. Candidate values $\lambda^*_i$ need to be screened, as not all cases
lead to valid solutions.
We discard negative $\lambda_i^*$ or smaller than the current $\lambda$;
those cases 
when the denominator $(\mathbf{S}^-\mathbf{C}^2 \mathbf{1})_i$ is zero; and 
those $\lambda_i^*$ that 
give negative  entries in the candidate solution 
\[\mathbf{C}^2\hat{\mathbf{u}}\biggr\rvert_{\lambda=\lambda^*_i}=
\mathbf{S}^-\mathbf{C}\mathbf{X}^T\mathbf{Y}
- \frac{(\mathbf{S}^-\mathbf{C}\mathbf{X}^T\mathbf{Y})_i}{(\mathbf{S}^-\mathbf{C}^2\mathbf{1})_i} 
\mathbf{S}^-\mathbf{C}^2 \mathbf{1}.\]
From candidates, we select the smallest positive $\lambda_i^*$ that gives non 
negative $\mathbf{C}^2\hat{\mathbf{u}}$.

\begin{example}\label{ex_mmp}
Continuing with Example \ref{exlarso}, we determine $\lambda$ at which the 
lasso path moves
from \texttt{--+} to a neighbor orthant.
Using Equation (\ref{ec_lambda}), we compute 
candidate $\lambda$ values 0.625, 0.2, -0.25 
that shrink coefficients $\beta_1,\beta_2,\beta_3$, respectively.
We screen candidates: the negative $\lambda$ is invalid so we are
left with the first two candidates. The first $\lambda$ 
gives $\mathbf{C}^2\hat{\mathbf{u}}$ with a 
negative entry and it is discarded and the second candidate gives non-negative
$\mathbf{C}^2\hat{\mathbf{u}}$ 
and it is selected. Thus at $\lambda=0.2$, the path moves 
from \texttt{--+} to the neighboring orthant 
 \texttt{-0+} by   shrinking $\beta_2$ to zero.
\end{example}

\subsubsection{Reactivation}\label{sec_reactivate}

The majority of lasso steps involves coefficient shrinkage and 
the lasso path is a series of shrinkages
while keeping track of increasing $\lambda$
and updating the orthant matrix $\mathbf{C}$. On occasion, a parameter that has been 
previously shrunk to zero becomes active, i.e. the path 
moves to a neighboring higher dimensional orthant. 

Reactivation can only take place when some entries in the
diagonal of $\mathbf{C}$ are zero. We 
reactivate by considering a new orthant matrix $\mathbf{C}'$, obtained from $\mathbf{C}$ 
by replacing 
a zero entry in the diagonal by $+1$ and checking if shrinking from $\mathbf{C}'$ 
gives a valid $\lambda$. 
This move is also done by changing the zero to $-1$ so for 
every zero in the diagonal of $\mathbf{C}$ we have two potential matrices $\mathbf{C}'$.
For every potential matrix $\mathbf{C}'$, computation of candidate $\lambda$ 
with formula (\ref{ec_lambda}) is done for all coordinates non zero entries 
in $\mathbf{C}'$. 
For a given $\mathbf{C}$, the number of neighboring 
orthants is $2(p-\mathbf{1}^T\mathbf{C}^2\mathbf{1})$,
i.e. twice the number of zero entries in the diagonal of $\mathbf{C}$.

\begin{example} Assume that the procedure is in 
orthant \texttt{0+-0}. To reactivate, we explore neighboring higher dimensional orthants 
 \texttt{++-0}, \texttt{-+-0}, \texttt{0+-+} and \texttt{0+--} and see if we can 
 reach \texttt{0+-0} by shrinking.
 As contrast, with direct shrink  from \texttt{0+-0}, we  see if the lasso path moves to 
 lower dimensional orthants \texttt{0+00} or \texttt{00-0}. 
\end{example}

Ultimately, reactivation is another shrinkage step. That is, 
moving from $\mathbf{C}$ to higher dimensional $\mathbf{C}'$ is equivalent to
shrinking from $\mathbf{C}'$ to $\mathbf{C}$. Of all computations with 
$\mathbf{C}'$ matrices, the smallest $\lambda$ with a valid 
 solution is selected. 

\subsection{Sequential lasso 2: the algorithm}\label{sec_theactualalgorithm}

We create the lasso path by sequentially using shrinkage and reactivation. We require
an algorithm for the shrinkage
step, which is used in the main algorithm.
We next describe both algorithms. 

Algorithm \ref{algostep} implements Equation (\ref{ec_lambda}), which
is the core of the lasso procedure. In this algorithm, $\mathbf{X}$ and $\mathbf{Y}$ are 
the same values used in the main algorithm.

Example \ref{ex_mmp} was built with \texttt{--+} for the
diagonal of $\mathbf{C}$  
  and calling
 Algorithm \ref{algostep} reiteratively with $i=1,2,3$. The computed 
  $\lambda$ values of the example are those 
that shrink each coordinate of $\beta$ to zero. The value $\lambda_c$ 
 is not given in the example so we could use $\lambda_c=0$ to guarantee valid 
 shrinkage moves. When in the lasso procedure, the value $\lambda_c$  is passed
 from the main algorithm to  Algorithm \ref{algostep}.

\begin{center}
\begin{algorithm}[H]
\SetAlgoLined\SetKwInput{KwData}{Input}\SetKwInput{KwResult}{Output}
\KwData{Orthant of interest $\mathbf{C}'$, index of coordinate to shrink
the path $i$ and $\lambda_c$ current value of parameter $\lambda$.}
\KwResult{Candidate values $\hat\lambda$, $\hat\beta$ and criterion $\hat L$ 
corresponding to critical 
change in orthant $\mathbf{C}'$ for $i$-th coordinate.}
 Compute inverse $\mathbf{S}^-=(\mathbf{C}'\mathbf{X}^T\mathbf{X}\mathbf{C}')^-$. 

	If ${(\mathbf{S}^-\mathbf{C}^2\mathbf{1})_i}\neq 0$, 
	compute 
	
		$\lambda^*_i={(\mathbf{S}^-\mathbf{C}'\mathbf{X}^T\mathbf{Y})_i}/{(\mathbf{S}^- \mathbf{C}^2\mathbf{1})_i}.$ \label{SE1}

With $\lambda^*_i$, compute 
$\mathbf{C}'^2\hat{\mathbf{u}}=\mathbf{S}^-\mathbf{C}'\mathbf{X}^T\mathbf{Y} - \lambda^*_i\mathbf{S}^-\mathbf{C}^2\mathbf{1}$, $\hat\beta=\mathbf{C}'\hat{\mathbf{u}}$
and $\hat L_{C'}$.

If ${(\mathbf{S}^-\mathbf{1})_i}= 0$, or if there are negative entries 
in $\mathbf{C}'^2\hat{\mathbf{u}}$, 
or if $\lambda^*_i\leq\lambda_c$ then 
 set output $RES:=\{\}$, otherwise set 
$RES:=\{\hat\lambda:=\lambda^*_i,\hat\beta,\hat L:=\hat L_{C'}\}.$

Return $RES$. 
 \caption{Shrinkage step for $i$-th coordinate}\label{algostep}
\end{algorithm}
\end{center}

Algorithm \ref{algo} is our main procedure. The algorithm builds the path
 from the ordinary least squares estimate and proceeds 
 by a series of shrinkage and reactivation movements. At a given 
step in the path, the algorithm
explores neighboring orthants and moves in the direction of steepest descent
determined by the smallest valid candidate $\lambda$ in step \ref{SF1}. Because
this move is in the lasso path, the quantity $L^*$ 
which was computed as $\hat L_\mathbf{C}$ is the minimal value of criterion $L$ 
at $\lambda=\lambda^*$.

Algorithm \ref{algo}  is guaranteed to always terminate because, in the worst case 
scenario, it will 
visit the complete list of all orthants, which is finite. In practice, the algorithm 
only visits a subset of all orthants.

\begin{center}
\begin{algorithm}[H]
\SetAlgoLined\SetKwInput{KwData}{Input}\SetKwInput{KwResult}{Output}
\KwData{Design model matrix $X$ with $p$ columns, vector of observations $Y$.}
\KwResult{Lasso path with $\lambda$, $\beta(\lambda)$ and $L$ at path breakpoints.}
 Initialization: Compute $\hat\beta=(\mathbf{X}^T\mathbf{X})^{-1}\mathbf{X}^T\mathbf{Y}$. 
Set matrix $\mathbf{C}:=\mbox{diag}(\sign(\hat\beta))$;
compute $\hat L_\mathbf{C}$; set
 $\lambda:=0$ and  output $O:=\{\{\ \lambda,\hat\beta,\hat L_\mathbf{C}\}\}$.

\While{$\mathbf{C}\neq \mathbf{0}$}{

\For{$j\in \{1,\ldots,p\}$}{\label{initioloop} 
\eIf{$C_{j,j}=0$}{
   \For{$k\in\{-1,1\}$}{
	
	Set $\mathbf{C}':=\mathbf{C}$ and update $C'_{j,j}:=k$.
	
	\For{$i\in\{1,\ldots,p\mbox{ such that }C'_{i,i}\neq 0\}$}{
	Run Algorithm \ref{algostep} with 
	inputs $\mathbf{C}',i,\lambda_c:=\lambda$. 
		Nonempty outputs $\{\hat\lambda$, $\hat\beta$ and $\hat L\}$ are kept
		until used in Step \ref{SF1}. \label{SE}
	}
	
	}

}{
  Set $\mathbf{C}':=\mathbf{C}$. 
  
  Run Algorithm \ref{algostep} with 
	inputs $C',i:=j,\lambda_c:=\lambda$. 
		Nonempty outputs $\{\hat\lambda$, $\hat\beta$ and $\hat L\}$ are kept
		until used in Step \ref{SF1}.\label{SC}
  
}

}\label{endloop}

From the set of all nonempty outputs $\{\{\hat\lambda,\hat\beta,\hat L\}\}$ of the loop
in steps \ref{initioloop}-\ref{endloop},
\label{SF1} 
select the smallest $\hat\lambda$. Call this 
$\lambda^*$, with associated $\beta^*, L^*$. 

Update output $O$ with these
values, i.e. $O:=O\cup \{\lambda^*,\beta^*,L^*\}$.

Update $\lambda:=\lambda^*$ and 
$\mathbf{C}:=\mbox{diag}(\sign(\beta^*))$.\label{SF2}

}

 \caption{Orthant lasso}\label{algo}
\end{algorithm}
\end{center}

\subsection{Detailed lasso example}\label{sec_example}

The lasso path we describe 
uses the data of Table \ref{tab_ex1d}(a) and 
was selected because it requires a reactivation 
step despite its small size. The path has initial shrinkage, reactivation of 
a variable and a final series of shrinkage steps. In Appendix 3 we detail the moves
of the algorithm as the path traverses through orthants.

Table \ref{tablepath}(a) summarizes the breakpoints of the lasso
path for this example. Each row lists $\lambda$, vector of coefficients $\beta$ and
criterion $L$ at a breakpoint of the path. 
The list of orthants involved in the path is \texttt{++-}, \texttt{0+-}, 
\texttt{-+-},  \texttt{-0-}, \texttt{-00}, \texttt{000}, which
can be seen from right to left in the standard plot of the lasso path of 
Figure \ref{fig_lasso_net}(a).
Table \ref{exexahustive} in Appendix 3 details all moves for this 
example. The table should be read from the top, as rejection
of candidates $\hat\lambda$ depends on the current value
of $\lambda$.

\begin{figure}
\begin{center}
\begin{tabular}{c}
\begin{knitrout}
\definecolor{shadecolor}{rgb}{0.969, 0.969, 0.969}\color{fgcolor}
\includegraphics[width=\maxwidth]{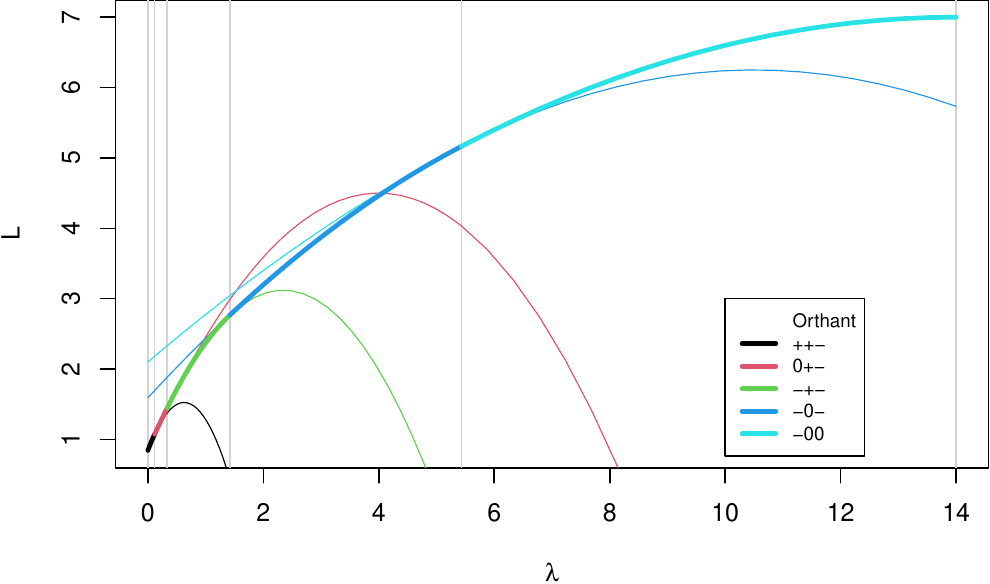} 
\end{knitrout}
\\
(a)\\\\
\begin{knitrout}
\definecolor{shadecolor}{rgb}{0.969, 0.969, 0.969}\color{fgcolor}
\includegraphics[width=\maxwidth]{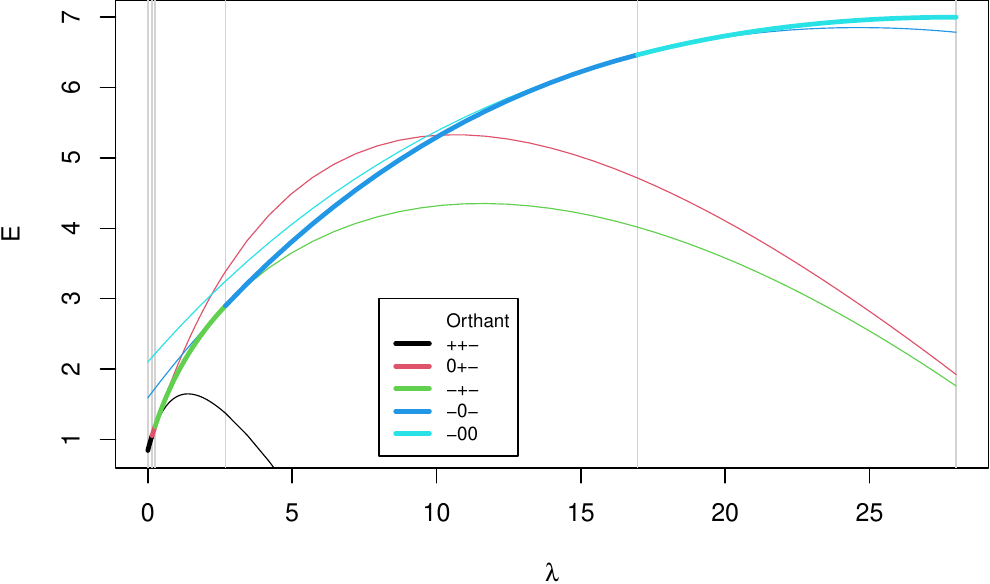} 
\end{knitrout}
\\
(b)
\end{tabular}
\end{center}
\caption{(a) Criterion $L$ of Lasso as the
path moves along orthants.
Panel (b) has $E$ of elastic net with $\alpha=0.5$. Both
cases use data in Table \ref{tab_ex1d}(a).}\label{fig_lasso_net_crit}
\end{figure}

Figure \ref{fig_lasso_net_crit}(a) shows  
  criterion $L$ as a function
 of $\lambda$ along the path, i.e. we
 plot $\hat L_\mathbf{C}$ for the orthants in the path.
 Colors indicate orthants, with
 bold line when the lasso path traverses along the orthant 
 and $\hat L_\mathbf{C}$ becomes $L$, and with
thin line when $\hat L_\mathbf{C}$ is not 
in the path. 
Finally, Figure \ref{fig_allorth} in the Appendix 
shows potential lasso moves after exhaustive
orthant exploration and rejection of unsuitable moves.

\begin{table}
\begin{center}
\begin{tabular}{cc}Lasso&Elastic net\\
$\begin{array}{l|rrr|l}\hline
\lambda&\multicolumn{3}{|c|}{\beta}&L\\\hline
0&0.114&0.871&-1.186&0.843\\
0.118&0&0.735&-1.029&1.074\\
0.333&0&0.667&-1&1.444\\
1.419&-0.372&0&-0.395&2.765\\
5.429&-0.429&0&0&5.163\\
14&0&0&0&7\\\hline
\end{array}$&
$\begin{array}{l|rrr|l}\hline
\lambda&\multicolumn{3}{|c|}{\beta}&E\\\hline
0&0.114&0.871&-1.186&0.843\\
0.146&0&0.732&-1.026&1.054\\
0.247&0&0.704&-1.013&1.181\\
2.687&-0.374&0&-0.359&2.898\\
16.961&-0.194&0&0&6.465\\
28&0&0&0&7\\\hline
\end{array}$\\
(a)&(b)\end{tabular}
\end{center}\caption{(a) Lasso path of the example of Section \ref{sec_example} and (b)
Elastic net path for the same data and $\alpha=0.5$, see Example \ref{ex_netfirst}.}\label{tablepath}
\end{table}

\begin{figure}
\begin{tabular}{cc}
\begin{knitrout}
\definecolor{shadecolor}{rgb}{0.969, 0.969, 0.969}\color{fgcolor}
\includegraphics[width=\maxwidth]{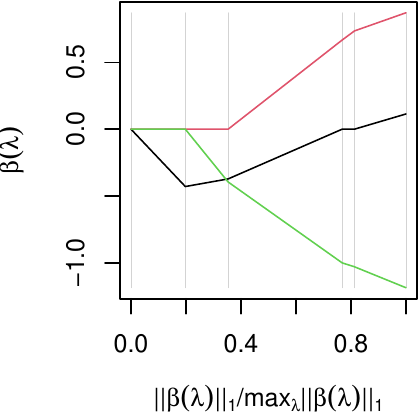} 
\end{knitrout}
&
\begin{knitrout}
\definecolor{shadecolor}{rgb}{0.969, 0.969, 0.969}\color{fgcolor}
\includegraphics[width=\maxwidth]{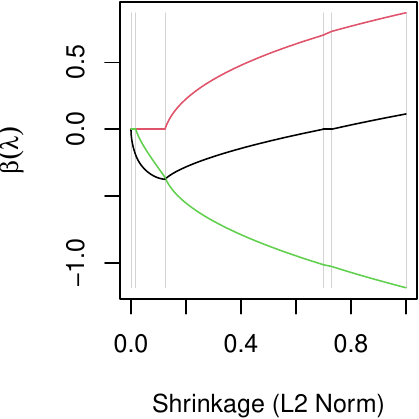} 
\end{knitrout}
\\
(a)&(b)\\\\
\begin{knitrout}
\definecolor{shadecolor}{rgb}{0.969, 0.969, 0.969}\color{fgcolor}
\includegraphics[width=\maxwidth]{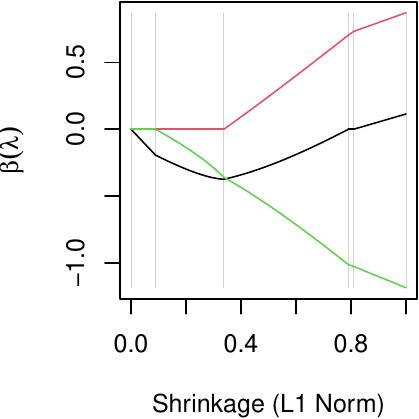} 
\end{knitrout}
&
\begin{knitrout}
\definecolor{shadecolor}{rgb}{0.969, 0.969, 0.969}\color{fgcolor}
\includegraphics[width=\maxwidth]{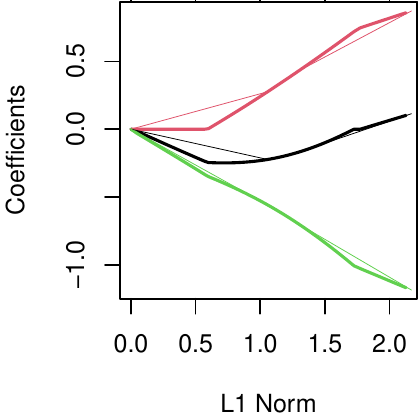} 
\end{knitrout}
\\
(c)&(d)\\
\end{tabular}
\caption{Shrinkage results of Table \ref{tab_ex1d}(a) data.
Panel (a) has Lasso of Table \ref{tablepath}(a); panels
(b) and (c) have the elastic net of Table \ref{tablepath}(b). Panel
(d) has \texttt{glmnet} results, further described in
 Example \ref{ex1glmnet}. 
The colors black, red, green in the plots
correspond to trajectories of coefficients $\beta_1,\beta_2,\beta_3$, respectively.
}\label{fig_lasso_net}
\end{figure}

\section{Elastic net}\label{sec_en}

Elastic net regularization combines model selection of lasso with improved prediction 
features of $L_2$ penalization.  The criterion to be minimized is 
$E$ of Equation (\ref{ecnet}) in Section \ref{sec_contrib}.
The nonnegative $\lambda$ controls the parameter penalization relative to
residual sum of squares; while $\alpha$ is a fixed number in $(0,1]$ that balances between 
the penalization  $||\beta||_1$ of lasso, achieved when $\alpha=1$ and
the quadratic penalty $||\beta||_2^2$ used in ridge regression and reached when $\alpha\to 0$. We exclude
$\alpha=0$ because at that point there is no shrinkage to zero 
for finite $\lambda$. 

Elastic net \cite{ZH2005} has been shown to improve over 
 lasso when predictors are heavily correlated \cite{FTH2010}. 
Lasso methods can be used in estimation of elastic net \cite{ZH2005},
and an implementation of the elastic net using coordinate descent is the \texttt{R} library
\texttt{glmnet}, see  \cite{FTH2010}.
A recent
 version of the
elastic net criterion uses s-estimators to improve estimation and variable selection performance under heavy tailed error 
distributions \cite{K2023}.
We next develop the orthant to estimation in elastic nets.

\subsection{Elastic net by orthants}

The orthant development for the elastic net mirrors what was done earlier for lasso, setting
$\beta=\mathbf{C}\mathbf{u}$ so that over the orthant determined by $\mathbf{C}$ the criterion is
\begin{equation}E_\mathbf{C}=\frac{1}{2}\mathbf{Y}^T\mathbf{Y}-
\mathbf{u}^T \mathbf{C}\mathbf{X}^T\mathbf{Y}
+\frac{1}{2}\mathbf{u}^T \mathbf{C}\mathbf{X}^T\mathbf{X}\mathbf{C}\mathbf{u}
+\lambda\alpha \mathbf{u}^T \mathbf{C}^2 \mathbf{1}
+ \lambda\frac{1-\alpha}{2}\mathbf{u}^T\mathbf{C}^2\mathbf{u}.\label{eenet}\end{equation}
The entries of vector $\mathbf{u}$ are required to be 
positive, although there is no mathematical restriction
for the entries of $\mathbf{u}$, which can be real numbers. In other words, $E_\mathbf{C}$ is a well formulated quadratic
form, which was also the case for $L_\mathbf{C}$.

The solution to the minimization of $E_\mathbf{C}$  
is the system 
\[\left(\mathbf{C}\mathbf{X}^T\mathbf{X}\mathbf{C}+\lambda(1-\alpha)\mathbf{C}^2\right)\mathbf{u}=\mathbf{C}\mathbf{X}^T\mathbf{Y}-\lambda\alpha \mathbf{C}^2\mathbf{1}.\] 
The matrix $\mathbf{C}\mathbf{X}^T\mathbf{X}\mathbf{C}+\lambda(1-\alpha)\mathbf{C}^2$ 
is the orthant counterpart of regularising
$\mathbf{X}^T\mathbf{X}$ with a multiple of the identity matrix in ridge regression, also known as
Tikhonov's regularization. The next theorem gives a 
property
of the generalized inverse of this matrix.
We omit its proof, which is similar to that of Theorem \ref{sinverse}.

\begin{theorem}\label{slinverse}
Let $\mathbf{S}(\lambda):=\mathbf{C}\mathbf{X}^T\mathbf{X}\mathbf{C}+\lambda(1-\alpha)\mathbf{C}^2$ and
let $\mathbf{S}(\lambda)^-$ be its generalized inverse.
Then $\mathbf{S}(\lambda)^-$ satisfies $\mathbf{S}(\lambda)\mathbf{S}(\lambda)^-=\mathbf{S}(\lambda)^-\mathbf{S}(\lambda)=\mathbf{C}^2$.
\end{theorem}

Using Theorem \ref{slinverse}, we have the solution 
\begin{equation}
\mathbf{C}^2\hat{\mathbf{u}}=\mathbf{S}(\lambda)^-\left(\mathbf{C}\mathbf{X}^T\mathbf{Y}-\alpha\lambda{}\mathbf{C}^2\mathbf{1}\right)\label{ec_c2unet}\end{equation}
and by using $\hat\beta=\mathbf{C}\hat{\mathbf{u}}$, retrieve
 the elastic net estimate
\begin{equation}\hat\beta=\mathbf{C}\mathbf{S}(\lambda)^-\mathbf{C}
\left(\mathbf{X}^T\mathbf{Y}-\alpha\lambda{}\mathbf{C}\mathbf{1}\right).\label{ecsolsnet}\end{equation}
The elastic net  
trajectory $\hat\beta$ starts from the ridge 
estimate $\mathbf{C}\mathbf{S}(\lambda)^-\mathbf{C}\mathbf{X}^T\mathbf{Y}$ and moves 
in the direction $-\alpha{}\lambda{}\mathbf{C}\mathbf{S}(\lambda)^-\mathbf{C}^2\mathbf{1}$.
This trajectory minimizes $E_\mathbf{C}$ over the orthant 
determined by $\mathbf{C}$, in other words, it is the exact elastic net path with no 
approximations involved. 
When formulated as a na\"ive elastic net, $\hat\beta$ of Equation (\ref{ecsolsnet}) 
coincides with estimator for orthonormal $\mathbf{X}$ of Equation (6) in \cite{ZH2005}. 

The notation $\mathbf{S}(\lambda)$ in Theorem \ref{slinverse} and elsewhere emphasizes 
the main role of $\lambda$:  
although  $\mathbf{S}(\lambda)$ depends on $\alpha$ and $\lambda$, in analyses $\alpha$ is kept fixed. 
We give an example of computation 
of $\mathbf{S}(\lambda)^-$ in Appendix 2.

Given the dependence of direction of descent on $\lambda$ through the 
matrix $\mathbf{S}(\lambda)^-$, the trajectories 
of net coefficients are not piecewise linear functions of $\lambda$ as with lasso. 
The following 
example shows nonlinearity of $\hat\beta$ even for a single explanatory variable. Example
\ref{ex_netpath} shows computation of $\hat\beta$ for a given orthant and range of $\lambda$.

\begin{example}
Consider data of Table \ref{tab_ex1d}(a) with single explanatory variable $\mathbf{X}_1$ 
as in Examples \ref{ex_1d3} and \ref{ex_1d3bis}.
 The 
least squares estimator $\hat\beta= -0.7$ lies 
in orthant \texttt{-} so $C=-1$. Using
$\mathbf{X}^T\mathbf{X}=20$, the generalized
inverse $\mathbf{S}(\lambda)^-$ is the 
scalar $\left(20+\lambda(1-\alpha)\right)^{-1}$,
and the net path 
is $\hat\beta=C\hat{\mathbf{u}}=
-1/\left(20+\lambda(1-\alpha)\right)\cdot \left(14-\alpha\lambda\right),$ 
where we used 
$\mathbf{X}^T\mathbf{Y}=-14$.
By substituting $\alpha=1$ in the net path $\hat\beta$, we retrieve 
the lasso estimate of Example \ref{ex_1d3}.
\end{example}

\begin{example}\label{ex_netpath}
In Figure \ref{fig_lapath} (right) we give part of the elastic 
net path for analysis of Table \ref{tablepath}(b). 
This segment is computed with Equation (\ref{ecsolsnet}) and 
orthant \texttt{-+-}, that corresponds to the path between 
rows $3$ and $4$ of the table. 
The trajectories are shown in solid line as they cut through \texttt{-+-}
for $\lambda\in (0.247, 2.687)$, and in dashed line outside the stated range of $\lambda$ at 
which point the trajectories are outside \texttt{-+-}. 
\end{example}

For given $\lambda$ and $\mathbf{C}$, by substituting $\hat\beta=\mathbf{C}\hat{\mathbf{u}}$ 
of Equation (\ref{ecsolsnet}) into $E_\mathbf{C}$ of
Equation (\ref{eenet}), we obtain the smallest value of elastic net criterion $E_\mathbf{C}$. 
This is a nonlinear function of $\lambda$ and $\alpha$ with the following expression
\begin{eqnarray}\hat E_\mathbf{C}=\frac{1}{2}\Big(&\mathbf{Y}^T\mathbf{Y} - \mathbf{Y}^T\mathbf{X}\mathbf{C}\mathbf{S}(\lambda)^-\mathbf{C}\mathbf{X}^T\mathbf{Y}-2\lambda\alpha\mathbf{1}^T\mathbf{S}(\lambda)^-\mathbf{C}\mathbf{X}^T\mathbf{Y}
-\lambda^2\alpha^2\mathbf{1}^T\mathbf{S}(\lambda)^-\mathbf{1}&\nonumber\\\nonumber
&+\lambda(1-\alpha)\mathbf{Y}^T\mathbf{X}\mathbf{C}\mathbf{S}(\lambda)^-\mathbf{S}(\lambda)^-\mathbf{C}\mathbf{X}^T\mathbf{Y}
-2\lambda^2\alpha\mathbf{1}^T\mathbf{S}(\lambda)^-\mathbf{S}(\lambda)^-\mathbf{C}\mathbf{X}^T\mathbf{Y}&\\\nonumber
&+\lambda^3\alpha^2(1-\alpha)+\mathbf{1}^T\mathbf{S}(\lambda)^-\mathbf{S}(\lambda)^-\mathbf{1}&     \Big).
\end{eqnarray}

\subsection{All orthants net analysis}

The all orthant approach of Section \ref{sec_orthant} can be applied with little
change to the elastic net, i.e.
for a given $\lambda$, consider all orthants and select the $\hat\beta$ that minimizes $\hat E_\mathbf{C}$. The same
screening considerations for orthant lasso must be used: discard those cases for 
which $\mathbf{C}^2\hat{\mathbf{u}}$ has 
negative entries, equivalently discard when $\hat\beta$ is not in orthant $\mathbf{C}$. 
All orthant computations can be done for a collection $\Lambda$ of values of $\lambda$ and
has exponential cost, akin to the situation 
described in Section \ref{sec_orthant}.

\subsection{Sequential approach to elastic net}\label{sec_netseq}

With a minor change, Algorithm \ref{algo} 
can be applied to the computation of the elastic net path. 
Consider an elastic net path in orthant $\mathbf{C}$. We find a breakpoint for
changing orthants for the
$i$-th coordinate by solving $(\mathbf{C}^2\hat{\mathbf{u}})_i=0$, i.e. 
\begin{equation}(\mathbf{S}(\lambda)^-\mathbf{C}\mathbf{X}^T\mathbf{Y})_i-\alpha\lambda({}\mathbf{S}(\lambda)^-\mathbf{C}^2\mathbf{1})_i=0,
\label{eq_enets}\end{equation}
which has to be solved for $\lambda$.
Rearranging this expression
leads to 
\begin{equation}\lambda_i^*=
\frac{(\mathbf{S}(\lambda_i^*)^-\mathbf{C}\mathbf{X}^T\mathbf{Y})_i}{\alpha( \mathbf{S}(\lambda_i^*)^-\mathbf{C}^2\mathbf{1})_i} \label{ec_net},\end{equation} 
which generalizes Equation (\ref{ec_lambda}) and 
depends on $\lambda^*_i$ on both sides. The modification
of Step \ref{SE1} in Algorithm \ref{algostep} is to solve numerically Equation (\ref{eq_enets}), that is

\noindent\textbf{3} Solve $(\mathbf{S}(\lambda)^-\mathbf{C}\mathbf{X}^T\mathbf{Y})_i-\alpha\lambda({}\mathbf{S}(\lambda)^-\mathbf{C}^2\mathbf{1})_i=0$
for $\lambda$ and call $\lambda_i^*$ to the solution.

We give two examples of elastic net computation, and 
in sections \ref{sec_solution} and \ref{sec_implementation}  we discuss
our implementation and compare against \texttt{glmnet}.

\begin{example}\label{ex_netfirst}
An elastic net with $\alpha=0.5$ was fitted to data of Example \ref{tab_ex1d}(a)
using Algorithms \ref{algostep} and \ref{algo} with the adaptation discussed 
above. 
Table \ref{tablepath}(b) shows the breakpoints at which the elastic net path changes orthant. 
The coefficients $\hat\beta$ are not piecewise linear
functions of $\lambda$, however they are computed easily using Equation (\ref{ecsolsnet}) with the appropriate orthant $C$. In Figure \ref{fig_lasso_net} panels (b) and (c) we show the elastic 
net path, and
in Figure \ref{fig_lasso_net_crit}(b) we show the evolution of criterion $E$ as it crosses 
 orthants of the elastic net path in its shrinking route towards zero.
\end{example}

\begin{example}
Figure \ref{fig_lasso_netex2} (Appendix) shows elastic net path fits for a synthetic dataset of 
$n=12$ observations in $p=10$ variables. Two values of $\alpha$ were used 
for the analysis and in both cases, the steps in the net trajectory were only shrinkage steps.
\end{example}

The criterion of elastic net can be studied  
plotting  $E$ against the shrinkage parameter $\lambda$. 
Figure \ref{fig_lasso_net_crit}(b) shows the evolution of 
piecewise nonlinear criterion $E$ for the data of Example \ref{ex_netfirst}. 
Line colors indicate orthants in the path,
with bold whenever the net path is
 traversing the orthant and thin line for suboptimal curves, that is
when the elastic net path is in another orthant.

\section{Conclusions and further discussion}\label{sec_final}

We have presented a new orthant method for the computation of 
lasso and elastic net estimates. 
Our proposal uses simple calculus techniques
and gives exact results. We proposed an algorithm to build the path
that avoids expensive
orthant evaluation and that has worked well in the examples we tried.
We briefly elaborate on issues still 
pending concerning implementation and theory development.

\subsection{Lasso computations and implementation}\label{sec_lassocomp}

The algorithm for lasso by orthants requires the iterative solution of
Equation (\ref{sols}) for the $i$-th component. This is a linear equation
whose explicit solution is Equation (\ref{ec_lambda}) that has proved to be remarkably stable.
A prototype \texttt{R} implementation \texttt{lassoq} of our algorithm for 
lasso path is given in Appendix 4 of this paper. 

Our code gives mostly the same results as the \texttt{lars} 
implementation of lasso \cite{HE2022}, although on some instances, it improves over
it as in the following example. 

\begin{example} For the data of Table \ref{tab_ex1d}(b) and 
for $0<\lambda<0.0833$, \texttt{lars} 
gives the lasso path as 
$(-1.25, -0.3333, 0.0833)^T-\lambda\cdot(0, 1, 1)^T$. 
This path makes the $L_1$ norm of
$\hat\beta$ a constant in \texttt{--+} and is not in the direction of 
maximum descent. The orthant computation 
with the same data and initial lasso step
of Example \ref{exlarso} gives a trajectory in the direction of maximum
descent.
 \end{example}

\subsection{Solving the orthant net equation}\label{sec_solution}

The elastic net path by orthants requires solving Equation (\ref{eq_enets}), 
i.e. finding $\lambda$ for which the $i$-th component
 of $\mathbf{C}^2\hat{\mathbf{u}}$ is zero. In essence, this is solving a univariate 
 non-linear equation and we survey classical approaches to this
 problem.
 
 Simple iteration of Equation (\ref{ec_net}) starts from initial $\lambda_0$ and 
 for $j=1,2,\ldots$ computes  
 $\lambda_j=
 {(\mathbf{S}(\lambda_{j-1})^-\mathbf{C}\mathbf{X}^T\mathbf{Y})_i}/{\left(\alpha( \mathbf{S}(\lambda_{j-1})^-\mathbf{C}^2 \mathbf{1})_i\right)}.$ 
 Another possibility 
 is Newton's method that iterates
$\lambda_j=\lambda_{j-1}- (\mathbf{C}^2\hat{\mathbf{u}})_i/(\mathbf{S}(\lambda_{j-1})^-((1-\alpha)\mathbf{C}^2\hat{\mathbf{u}}-\alpha\mathbf{1})  )_i$,
where $\mathbf{C}^2\hat{\mathbf{u}}$ also depends on $\lambda_{j-1}$.
In either approach, iteration continues until
 the absolute difference between  
 values $|\lambda_{j}-\lambda_{j-1}|$ is within a specified threshold.
  In our experience, these two iteration methods can lead in some cases, to 
  $\lambda_j$ oscillating outside the allowable range 
  $[0,\max_i|(X^TY)_i|/\alpha]$ and have not pursued its use.
 
We have experimented with two other iterative methods that have proved 
more  
stable in our numerical examples. One is the secant method, that does not 
require derivative
 information, and the other is the bisection method.
The latter method, although it is not 
the most efficient, is the method that has worked best for the orthant net 
method. 
We are exploring ways to improve the numerical stability and accuracy of numerical
solvers. This is 
work in progress, for which we have a prototype \texttt{R} implementation
\texttt{elastiq} given in Appendix 5 of this paper.

\subsection{Implementation of elastic net}\label{sec_implementation}

We look at results obtained with the orthant net
method and those of the \texttt{R} package \texttt{glmnet}.
This function regularizes generalized
linear models, and we use the option \texttt{family=gaussian}. 
Our comparison is not about the speed of computations nor about the dimensionality
of data, but about the quality of results obtained.

We emphasize that results obtained with the orthant net method 
are not approximations but are exact solutions to the minimization of 
criterion $E$.  
In contrast, \texttt{glmnet} appears designed not for precise results
but for fast approximate analysis. We compare net methods through examples and 
 discuss a simulation.

\begin{example}\label{ex1glmnet}
Two \texttt{glmnet} analyses were carried out with 
the data of Table \ref{tab_ex1d}(a) and
$\alpha=0.5$: one analysis without supplying $\lambda$ values
and another uses $\lambda$ breakpoints from the orthant net  
of Table\ref{tablepath}(b).
Both results are shown in Figure \ref{fig_lasso_net}(d), where  
bold lines are for the first analysis 
and thin lines are 
for the second analysis.

We compare results with orthant net results
of Figure \ref{fig_lasso_net}(c).
When $\lambda$ is not supplied in \texttt{glmnet}, the  patterns
of $\beta_1,\beta_2$ are similar to
the exact orthant values. However, the shrinkage pattern of $\beta_3$ is not 
correctly recovered and
this parameter shrinks to zero later than should be.
When $\lambda$ breakpoints are supplied, 
 \texttt{glmnet} analysis produces trajectories that differ substantially from the 
 orthant solution, with $\beta_2,\beta_3$ shrinking later than needed.
It is also surprising that
even the least squares estimate from \texttt{glmnet} for both cases only agrees with the 
correct value when rounded to a single digit.

As for orthants covered, \texttt{glmnet} generally recovers less orthants than the correct orthant solution. The elastic net orthant solution of Table \ref{tablepath}(b) goes through  orthants 
 \texttt{++-}, \texttt{0+-}, \texttt{-+-}, \texttt{-0-}, \texttt{-00}, \texttt{000}.
Without specifying $\lambda$, the \texttt{glmnet} path traverses through less orthants 
 \texttt{++-}, \texttt{0+-}, \texttt{-+-}, \texttt{-0-}, \texttt{000}, and finally, when specifying $\lambda$ breakpoints, the  \texttt{glmnet} path only crosses the
 orthants \texttt{++-}, \texttt{-+-}, \texttt{000}.
\end{example}

\begin{figure}[h]
\begin{tabular}{cc}
\begin{knitrout}
\definecolor{shadecolor}{rgb}{0.969, 0.969, 0.969}\color{fgcolor}
\includegraphics[width=\maxwidth]{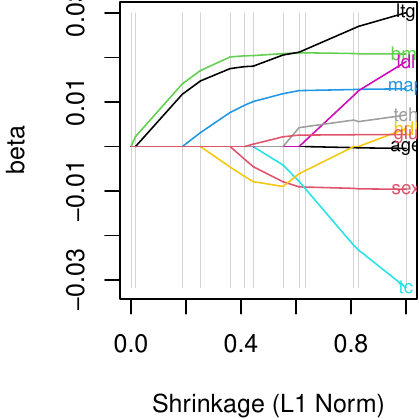} 
\end{knitrout}
&
\begin{knitrout}
\definecolor{shadecolor}{rgb}{0.969, 0.969, 0.969}\color{fgcolor}
\includegraphics[width=\maxwidth]{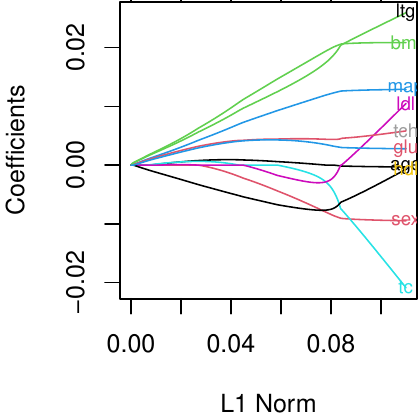} 
\end{knitrout}
\\
(a)&(b)\end{tabular}
\caption{Analysis of scaled diabetes data with (a) the elastic net orthant of this paper 
and (b) \texttt{glmnet}.}\label{fig_lasso_netex}
\end{figure}

\begin{example}\label{ex_quadscoincidence}
We carried out elastic net orthant and \texttt{glmnet} analyses for synthetic
data. Both analyses used $\alpha=0.5$ and \texttt{glmnet} was used without 
specifying $\lambda$. Default \texttt{glmnet} analysis does
$93$ steps with redundancy as they 
only cover $40$ orthants,
while the net orthant method yields $56$
steps with no redundant orthants retrieved. The orthants visited with
each methodology are given in Figure \ref{fig_quads} (Appendix).  
The \texttt{glmnet} agrees with 
$55.4\%$ of the orthants of
the net orthant approach.
\end{example}

\begin{figure}[h]
\begin{center}
\begin{knitrout}
\definecolor{shadecolor}{rgb}{0.969, 0.969, 0.969}\color{fgcolor}
\includegraphics[width=\maxwidth]{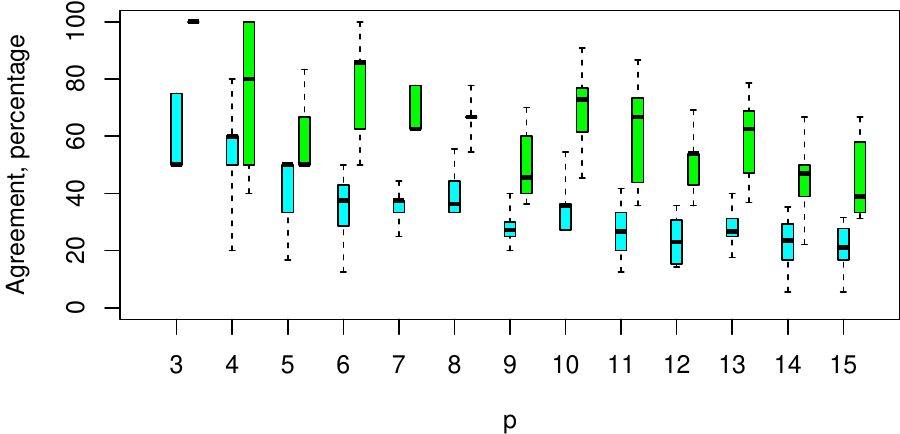} 
\end{knitrout}
\end{center}
\caption{Agreement between orthant method and \texttt{glmnet} for simulated data. In green,
agreement between orthant and \texttt{glmnet} analysis without specifying $\lambda$; 
in light blue, 
agreement when \texttt{glmnet} uses $\lambda$ provided by the orthant. 
The horizontal axis is the dimension $p$.
}\label{fig_wiwo}
\end{figure}

\begin{figure}
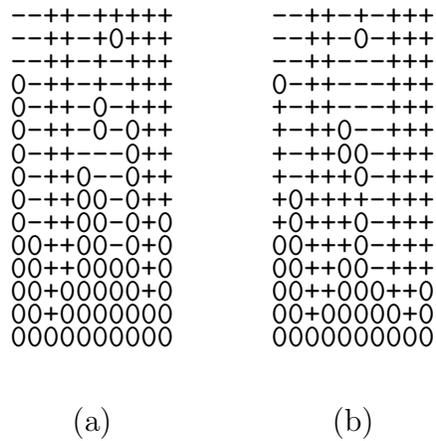

\begin{center}
\begin{tabular}{cc}
\begin{minipage}[t]{1.2in}
\begin{spacing}{0.6}\begin{center}
\texttt{--++-+++++\\ --++-+0+++\\ --++-+-+++\\ 0-++-+-+++\\ 0-++-0-+++\\ 0-++-0-0++\\ 0-++---0++\\ 0-++0--0++\\ 0-++00-0++\\ 0-++00-0+0\\ 00++00-0+0\\ 00++0000+0\\ 00+00000+0\\ 00+0000000\\ 0000000000}
\end{center}
\end{spacing}
\end{minipage}&
\begin{minipage}[t]{1.2in}
\begin{spacing}{0.6}\begin{center}
\texttt{--++-+-+++\\ --++-0-+++\\ --++---+++\\ 0-++---+++\\ +-++---+++\\ +-++0--+++\\ +-++00-+++\\ +-+++0-+++\\ +0++++-+++\\ +0+++0-+++\\ 00+++0-+++\\ 00++00-+++\\ 00++000++0\\ 00+00000+0\\ 0000000000}
\end{center}
\end{spacing}
\end{minipage}\\(a)&(b)
\end{tabular}
\end{center}

\caption{Orthant excursion for the scaled diabetes data set, computed with (a) the 
orthant method and (b) \texttt{glmnet}.}\label{tab_quads}
\end{figure}

\begin{example}
An elastic net analysis of the diabetes data set \cite{T1996} with $\alpha=0.5$
was performed. Two methods were used: elastic
net orthant and \texttt{glmnet} without specifying $\lambda$. 
The data for both analyses was scaled by
a constant $k=25000$. This scaling was used to stabilize the elastic net orthant 
computation. 
Figure \ref{fig_lasso_netex} shows the paths for both analyses, 
and despite some similarities, \texttt{glmnet} tends to collapse most trajectories at a single step
of the trajectory, contrary to the orthant method in which the trajectories 
shrink to zero at different points in the path. Figure \ref{tab_quads} (Appendix) shows 
orthant traverses for each method. The \texttt{glmnet} path has an agreement 
of only $20\%$ with the orthant method.
\end{example}

The examples suggest that \texttt{glmnet} results may 
differ with the orthant net method and we 
compared with simulation experiment. 
We simulated data with dimensions $p$ from $3$ 
to $15$, for each dimension simulating $42$ data sets.
For each set we fitted elastic net
using the orthant method as well as \texttt{glmnet},
using $\alpha$ from $0.3$ 
to $0.9$.
We recorded the orthants visited by every method, and 
computed the percentage of \texttt{glmnet} orthants
that agree with the orthant net method. Two versions of 
 \texttt{glmnet} were compared: 
without providing $\lambda$ and using breakpoints from 
the orthant net method. 
The agreement decreases with increasing $p$; and the agreement is lower 
when we provide breakpoint values of $\lambda$, 
see Figure \ref{fig_wiwo}. There is not much change of agreement as 
function of $\alpha$ (plot not shown).

\subsection{Further work}\label{sec_furtherwork}

Orthant work can be extended in several directions. 
Firstly, expectation of Equation (\ref{ec_beta})
for lasso or (\ref{ecsolsnet}) for elastic net allows exact
bias computation which could be used to compare results under
model misspecification. This analysis is contingent on choice of $\lambda$
and $\mathbf{C}$, and theory developments should consider ways of removing this dependence
from bias results.

A second line of work is the comparison of our algorithm with the
modified LARS algorithm \cite{EHJT2004} used to compute lasso.
By construction, our proposal already gives the optimal path, but we still have to
study the equivalence against the LARS algorithm, which we have already pointed that
has substandard performance in some instances. 
A related development is the removal 
of the reactivation step in Algorithm \ref{algo}. This would make orthant 
computations much faster and should be compared with the unmodified LARS algorithm.

A modified approach to lasso 
considers constraints on parameters to 
achieve polynomial hierarchy along the path \cite{ML2020}, generalizing  
hierarchical lasso work by \cite{BTT2013}.  The constrained approach to the 
path is carried out by numerical minimization over cones, with no apparent 
closed formul\ae{} available. 
Two possibilities arise here. One is the development over the cones themselves,
while would be to project the path
into the constrained region to create an approximate constrained path which would be
compared with the correct lasso or elastic net paths. 

Orthant methods can be adapted to
different ways of regularizing. A direct
case is the na\"ive elastic net with
penalty $\lambda_1||\beta||_1+\lambda_2||\beta||_2^2/2$,
 implemented in \texttt{LassoNet}, see \cite{WZZ2006,WSS2018}; or
 adaptive lasso \cite{WL2007} in which coefficients are weighed 
 in the penalty function  $\sum_{j=1}^pw_j|\beta_j|$. In both cases the
orthant approach can be used, see 
an  example of adaptive lasso in Appendix 6. However,  
  fused lasso \cite{TSRZK2005}
 with penalty  
 $\lambda_1||\beta||_1+\lambda_2\sum_{j=2}^n|\beta_j-\beta_{j-1}|$ 
 and the penalty $\lambda||\mathbf{D}\beta||_1$ of 
 generalized lasso \cite{TT2011}  would 
require careful consideration. These two latter cases are still 
 quadratic forms, not defined over
orthants, but rather over polyhedral cones. The computation of $\lambda$ breakpoints
should consider entry and exit points of paths between cones. 

\section*{Acknowledgements}

The author acknowledges partial funding by EPSRC travel grant
EP/K036106/1.

\bibliographystyle{unsrt} 
\bibliography{references}

\pagebreak

\section*{Appendixes}

\subsection*{Appendix 1 - Proof of Theorem \ref{sinverse}}

\begin{proof}
The proof is by construction. Without lack of generality, we assume that all non-zero entries 
in the diagonal of $\mathbf{C}$ take value one, and to avoid a trivial case, there is at least one 
non-zero entry in the diagonal of $\mathbf{C}$.

The matrix $\mathbf{X}\mathbf{C}$ has the same size as $\mathbf{X}$, but with some columns of $\mathbf{X}$ replaced by zero 
columns. The matrix $\mathbf{S}=\mathbf{C}\mathbf{X}^T\mathbf{X}\mathbf{C}$ is the same size as $\mathbf{X}^T\mathbf{X}$ and its contents are
equal those of $\mathbf{X}^T\mathbf{X}$ except for some zero rows and columns.
The location of the zero columns of $\mathbf{X}\mathbf{C}$ and the zero rows and columns of $\mathbf{S}$  corresponds to the
zeroes in the diagonal of $\mathbf{C}$.
The rank of $\mathbf{X}\mathbf{C}$ equals the number of non-zero entries in the diagonal of $\mathbf{C}$, because
the non-zero columns of $\mathbf{X}\mathbf{C}$ are linearly independent. The non-zero submatrix of $\mathbf{C}\mathbf{X}^T\mathbf{X}\mathbf{C}$ has also
the same rank as $\mathbf{X}\mathbf{C}$ and is invertible. The inverse $\mathbf{S}^-$ is the ordinary matrix 
inverse of the non-zero submatrix of $\mathbf{S}$, located according to non-zero entries in the 
diagonal of $\mathbf{C}$.

If $\mathbf{C}$ is full rank, then $\mathbf{C}=\mathbf{I}$ and the inverse 
is $\mathbf{S}^-=\mathbf{S}^{-1}$ so that 
$\mathbf{S}\mathbf{S}^-=\mathbf{I}=\mathbf{C}^2$. In general, when we multiply $\mathbf{S}\mathbf{S}^-$, we are doing the product of the smaller, nonzero invertible
 submatrix of $\mathbf{C}\mathbf{X}^T\mathbf{X}\mathbf{C}$ with its inverse, hence we always obtain an submatrix of the identity which
 is precisely $\mathbf{C}^2$, that is $\mathbf{S}\mathbf{S}^-=\mathbf{C}^2$ which has ones in the positions of non-zero entries in the diagonal of $\mathbf{C}$.
A similar argument is used to show that $\mathbf{S}^-\mathbf{S}=\mathbf{C}^2$.

In the case of non-zero entries of $\mathbf{C}$ taking value $-1$, the development described holds  
 because the rank of both $\mathbf{X}\mathbf{C}$ and $\mathbf{S}$ is not altered by some columns of $\mathbf{X}\mathbf{C}$  
 reversing sign and for some sign changes in columns and rows of $\mathbf{S}$.

The construction gives a unique matrix $\mathbf{S}^-$ which is a Moore-Penrose 
inverse as it satisfies $\mathbf{S}\mathbf{S}^-\mathbf{S}=\mathbf{S}$, $\mathbf{S}^-\mathbf{S}\mathbf{S}^-=\mathbf{S}^-$ and both  $\mathbf{S}^-\mathbf{S}$ and $\mathbf{S}\mathbf{S}^-$ are diagonal.
\end{proof}

Theorem \ref{sinverse} is still valid for the 
trivial case in the last step of lasso when $\mathbf{C}$ has all zeroes in its diagonal in 
which case $\mathbf{S}^-=\mathbf{C}$. The proof for Theorem \ref{slinverse} follows the rationale above as
$\mathbf{S}(\lambda)=\mathbf{C}\mathbf{X}^T\mathbf{X}\mathbf{C}+\lambda(1-\alpha)\mathbf{C}^2$ involves a submatrix of $\mathbf{X}^T\mathbf{X}$ regularized with 
a multiple of identity with corresponding dimensions.

\subsection*{Appendix 2 - Examples of $\mathbf{S}^-$ and $\mathbf{S}(\lambda)^-$}

We give one example of the computation of the inverse $\mathbf{S}^-$ and another
of $\mathbf{S}(\lambda)^-$. Both examples use the matrix $\mathbf{X}$ of 
Table \ref{tab_ex1d}(a).

\begin{example}
Consider the orthant \texttt{0+-}, i.e.
the matrix $\mathbf{C}$ has diagonal entries $0, 1, -1$ and
for computations we only use the columns $2, 3$ 
of $\mathbf{X}$. The inverse $\mathbf{S}^-$ is built with the usual inverse 
of the lower $2\times 2$ block. We have 
\[\mathbf{S}=\left(\begin{array}{rrr}
0&0&0\\
0&4&-2\\
0&-2&12\\
\end{array}\right),\;\;\mathbf{S}^-=\frac{1}{44}\left(\begin{array}{rrr}
0&0&0\\
0&12&2\\
0&2&4\\
\end{array}\right) \mbox{ and }\mathbf{S}\mathbf{S}^-=
\left(\begin{array}{rrr}
0&0&0\\
0&1&0\\
0&0&1\\
\end{array}\right).\]
\end{example}

\begin{example}
Consider $\mathbf{C}$ of orthant \texttt{-0-}. For 
$\lambda=10$ and $\alpha=0.5$ we have 
\[\mathbf{S}(\lambda)=\left(\begin{array}{rrr}
25&0&13\\
0&0&0\\
13&0&17\\
\end{array}\right),\;\;\mathbf{S}(\lambda)^-=\frac{1}{256}\left(\begin{array}{rrr}
17&0&-13\\
0&0&0\\
-13&0&25\\
\end{array}\right)\mbox{ and }\mathbf{S}(\lambda)\mathbf{S}(\lambda)^-=\left(\begin{array}{rrr}
1&0&0\\
0&0&0\\
0&0&1\\
\end{array}\right).\]
\end{example}

\subsection*{Appendix 3 - Orthant moves for Section \ref{sec_example}}

We detail the algorithm moves for the data of Table \ref{tab_ex1d}(a).

\begin{enumerate}
\item Initialization

The least squares estimate is $\hat\beta=(0.114, 0.871, -1.186)^T$.
We have orthant \texttt{++-}; 
compute $L=0.843$ and set $\lambda=0$.
 
\item Current orthant \texttt{++-}\label{run1}
with $\lambda=0$

Matrix $\mathbf{C}$ has no zero elements in its diagonal 
so no reactivation is done. We proceed to
 shrink every coordinate using Algorithm \ref{algostep} with $i=1,2,3$. 

Of the candidate $\hat\lambda$,
only $\lambda^*=0.118$ is valid and we have
$\beta^*=(0, 0.735, -1.029)^T$;
 criterion $L=1.074$ and
 update the orthant to \texttt{0+-}.

\item Current orthant  \texttt{0+-} \label{run2}
with $\lambda=0.118$

The diagonal of $\mathbf{C}$ has a zero and we reactivate,
i.e. substitute $C_{1,1}$ with each of $\mp 1$. 
Using $-1$ leads to orthant
\texttt{-+-}, and shrinking from this
orthant gives two valid candidates $\hat\lambda$.
Reactivating with $+1$ creates orthant \texttt{++-},
and shrinking from 
\texttt{++-} repeats the computation 
of step \ref{run1} above, only this time there are no valid $\hat\lambda$ candidates
because of the current value $\lambda=0.118$.

Shrinking 
from  \texttt{0+-} with $i=2,3$ gives two $\hat\lambda$, 
of which only one is valid.

We have three valid $\hat\lambda$ candidates. We select $\lambda^*=0.333$ with 
$\beta^*=(0, 0.667, -1)^T$ and criterion $L=1.444$. We remain 
in orthant \texttt{0+-}
because
of $\beta^*$.

\item Current orthant \texttt{0+-}
with $\lambda=0.333$

The moves are a second pass of what was already done in step \ref{run2}:
 reactivation to \texttt{-+-} and \texttt{++-}; shrinkage from \texttt{0+-}. 
Given the current value of $\lambda$, one earlier candidate from step \ref{run2} 
is not valid and we have two valid $\hat\lambda$ candidates.

We select $\lambda^*=1.419$ with parameter vector
$\beta^*=(-0.372, 0, -0.395)^T$, criterion $L=2.765$ and 
update orthant to 
\texttt{-0-}.

\item Current orthant \texttt{-0-}\label{run3}
with $\lambda=1.419$

The orthant has a zero in the second position so we carry a reactivation step,
 i.e. substituting $C_{2,2}$ with each of $\mp 1$, then shrink.
When reactivating the second entry with $-1$, we shrink from \texttt{---}. None of the three candidate $\hat\lambda$ values are suitable. We reactivate with $+1$ to \texttt{-+-}
and here we repeat part of step \ref{run2}. Given the current value $\lambda$, none of the 
candidate $\hat\lambda$ are valid.

After reactivation, we shrink from \texttt{-0-}. Of two
$\hat\lambda$, only one is valid.

We have a single candidate $\hat\lambda$ which we select: $\lambda^*=5.429$ with 
$\beta^*=(-0.429, 0, 0)^T$, criterion $L=5.163$ and updated orthant \texttt{-00}.

\item Current orthant \texttt{-00} with $\lambda=5.429$

Orthant \texttt{-00} has two 
zeroes and thus the reactivation step will explore
four orthants obtained by substituting $\mp 1$ in each of the positions $2$ and $3$.

Reactivating to \texttt{--0} leads to no valid $\hat\lambda$ candidates. We
have a similar situation when reactivating to 
\texttt{-+0} and at this point we have no valid $\hat\lambda$ candidates.

Reactivating to \texttt{-0-} and shrinking repeats part of step \ref{run3}, 
with no valid candidates given current $\lambda$.  
Reactivation to \texttt{-0+}  does not give valid candidates.

We do the only shrinkage move left from \texttt{-000}. This gives a valid
 $\hat\lambda$ that we select so $\lambda^*=14$ 
with 
$\beta^*=(0, 0, 0)^T$ and criterion value $L=7$. 
At this point the diagonal of $C$ is the zero vector \texttt{000}
and the procedure ends. 

\end{enumerate}

Table \ref{exexahustive} details all moves of Algorithm \ref{algo} for the
example in Section \ref{sec_example}. 
Recall that both moves R(eactivate) and S(hrink) 
use Algorithm \ref{algostep} for shrinkage, and the third column of the table 
gives the index $i$ used in every local shrinkage call
of Algorithm \ref{algostep} 
inside  Algorithm \ref{algo}. Note revisited orthants in the table, suggesting ways
to improve the algorithm and provided \texttt{R} code.

\begin{figure}
\begin{center}
\begin{knitrout}
\definecolor{shadecolor}{rgb}{0.969, 0.969, 0.969}\color{fgcolor}
\includegraphics[width=\maxwidth]{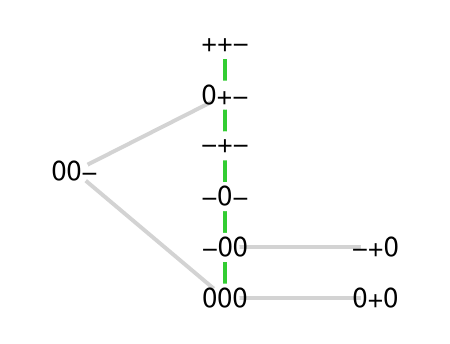} 
\end{knitrout}
\end{center}
\caption{Potential moves over all orthants:
lasso (green) and moves not minimizing $L$ (grey).
We only show valid moves 
with $\mathbf{C}^2\hat{\mathbf{u}}\geq\mathbf{0}$ and $\lambda\geq 0$.}\label{fig_allorth}
\end{figure}

Figure \ref{fig_allorth} shows nine potential moves for this data, when searching
over all $54$ orthant moves with Equation (\ref{ec_lambda}) and screening valid moves only. 
Besides five 
lasso moves, two other moves also appear in Table \ref{exexahustive}, 
while another two arise only with
exhaustive orthant search. 
Many invalid orthant moves are excluded from the figure, for 
example from \texttt{-+0} shrinking the first coordinate to \texttt{0+0}. 

\begin{table}
\begin{tabular}{|l|lr|c|l|}\hline
Current $\mathbf{C}$, $\lambda$&\multicolumn{2} {|c|} {Move and candidate $\hat\lambda$}&$i$&Comment\\\hline\hline
\multirow{3}{*}{\texttt{++-}, 0}&S from \texttt{++-} to \texttt{0+-}&$0.118$&$1$&Accepted move\\
&S from \texttt{++-} to \texttt{+0-}&$0.753$&$2$&Reject, non positive $\mathbf{C}^2\hat{\mathbf{u}}$\\
&S from \texttt{++-} to \texttt{++0}&$0.893$&$3$&Reject, non positive $\mathbf{C}^2\hat{\mathbf{u}}$\\\hline\hline
\multirow{8}{*}{\texttt{0+-}, 0.118}&R to \texttt{-+-} then \texttt{0+-} &$0.333$&$1$&Accepted move\\
&R to \texttt{-+-} then \texttt{-0-} &$1.419$&$2$&Reject, valid move but not minimum for $\hat\lambda$\\
&R to \texttt{-+-} then \texttt{-+0} &$2.128$&$3$&Reject, non positive $\mathbf{C}^2\hat{\mathbf{u}}$\\\cline{2-5}
&R to \texttt{++-} then \texttt{0+-}&$0.118$&$1$&Reject, $\hat\lambda\leq\lambda$ (this was an earlier step)\\
&R to \texttt{++-} then \texttt{+0-}&$0.753$&$2$&Reject, non positive $\mathbf{C}^2\hat{\mathbf{u}}$\\
&R to \texttt{++-} then \texttt{++0}&$0.893$&$3$&Reject, non positive $\mathbf{C}^2\hat{\mathbf{u}}$\\\cline{2-5}
&S from \texttt{0+-} to \texttt{00-}&$2.426$&$2$&Reject, valid move but not minimum $\hat\lambda$\\
&S from \texttt{0+-} to \texttt{0+0}&$7.666$&$3$&Reject, non positive $\mathbf{C}^2\hat{\mathbf{u}}$\\\hline\hline
\multirow{8}{*}{\texttt{0+-}, 0.333}&R to \texttt{-+-} then \texttt{0+-} &$0.333$&$1$&Reject, $\lambda\leq\lambda_c$ (this was an earlier step)\\
&R to \texttt{-+-} then \texttt{-0-} &$1.419$&$2$&Accepted move\\
&R to \texttt{-+-} then \texttt{-+0} &$2.128$&$3$&Reject, non positive $\mathbf{C}^2\hat{\mathbf{u}}$\\\cline{2-5}
&R to \texttt{++-} then \texttt{0+-}&$0.118$&$1$&Reject, $\hat\lambda\leq\lambda$ (this was an earlier step)\\
&R to \texttt{++-} then \texttt{+0-}&$0.753$&$2$&Reject, non positive $\mathbf{C}^2\hat{\mathbf{u}}$\\ 
&R to \texttt{++-} then \texttt{++0}&$0.893$&$3$&Reject, non positive $\mathbf{C}^2\hat{\mathbf{u}}$\\\cline{2-5}
&S from \texttt{0+-} to \texttt{00-}&$2.426$&$2$&Reject, valid move but not minimum for $\hat\lambda$\\
&S from \texttt{0+-} to \texttt{0+0}&$7.666$&$3$&Reject, non positive $\mathbf{C}^2\hat{\mathbf{u}}$\\\hline\hline
\multirow{8}{*}{\texttt{-0-}, 1.419}&R to \texttt{---} then \texttt{0--}&$-0.571$&$1$&Reject, non positive $\mathbf{C}^2\hat{\mathbf{u}}$\\
&R to \texttt{---} then \texttt{-0-}&$-2.179$&$2$&Reject, $\hat\lambda\leq\lambda$ \\
&R to \texttt{---} then \texttt{--0}&$-5.929$&$3$&Reject, $\hat\lambda\leq\lambda$ \\\cline{2-5}
&R to \texttt{-+-} then \texttt{0+-}&$0.333$&$1$&Reject, $\hat\lambda\leq\lambda$  (this was an earlier step)\\
&R to \texttt{-+-} then \texttt{-0-}&$1.419$&$2$&Reject, $\hat\lambda\leq\lambda$  (this was an earlier step)\\
&R to \texttt{-+-} then \texttt{-+0}&$2.128$&$3$&Reject, non positive $\mathbf{C}^2\hat{\mathbf{u}}$\\\cline{2-5}
&S from \texttt{-0-} to \texttt{00-}&$-25.000$&$1$&Reject, $\hat\lambda\leq\lambda$ \\
&S from \texttt{-0-} to \texttt{-00}&$5.429$&$3$&Accepted move\\\hline\hline
\multirow{9}{*}{\texttt{-00}, 5.429}&R to \texttt{--0} then \texttt{0-0}&$11.000$&$1$&Reject, non positive $\mathbf{C}^2\hat{\mathbf{u}}$\\
&R to \texttt{--0} then \texttt{-00}&$-0.286$&$2$&Reject, $\hat\lambda\leq\lambda$\\\cline{2-5}
&R to \texttt{-+0} then \texttt{0+0}&$18.333$&$1$&Reject, non positive $\mathbf{C}^2\hat{\mathbf{u}}$\\
&R to \texttt{-+0} then \texttt{-00}&$0.316$&$2$&Reject, $\hat\lambda\leq\lambda$\\\cline{2-5}
&R to \texttt{-0-} then \texttt{00-}&$-25.000$&$1$&Reject, $\hat\lambda\leq\lambda$\\
&R to \texttt{-0-} then \texttt{-00}&$5.429$&$3$&Reject, $\hat\lambda\leq\lambda$  (this was an earlier step)\\\cline{2-5}
&R to \texttt{-0+} then \texttt{00+}&$1.000$&$1$&Reject, non positive $\mathbf{C}^2\hat{\mathbf{u}}$\\
&R to \texttt{-0+} then \texttt{-00}&$-1.151$&$3$&Reject, $\hat\lambda\leq\lambda$\\\cline{2-5}
&S from \texttt{-00} to \texttt{000}&$14.000$&$1$&Accepted move, end of path\\\hline
\end{tabular}
\caption{Lasso computations for the example 
of Section 
\ref{sec_example}.}\label{exexahustive}
\end{table}

\pagebreak

\subsection*{Appendix 4 - Lasso \texttt{R} code and example}

Four functions are used: 
\texttt{lassoq} is Algorithm \ref{algo};  \texttt{shrink} is
Algorithm \ref{algostep}, with the crucial step \ref{SE1} 
that implements Equation (\ref{ec_lambda}) in the line 
\texttt{SMCXTY/SM1};
\texttt{pseudo} does $\mathbf{S}^-$ of Theorem \ref{sinverse} and \texttt{Lhat}
evaluates $\hat L_\mathbf{C}$ of Equation (\ref{ec_criterionopt}). 

The code is provided without guarantee. We do not 
accept responsibility for the accuracy of results nor for use or misuse of the 
code or results from it.

{\scriptsize
\begin{verbatim}
## Pseudoinverse of S=CX^TXC, with C a diagonal of {+-1,0} entries
pseudo<-function(XM,CM){  ## Define S, result SM, nonzero indices and invertible part of S
  S<-CM%*%t(XM)%*%XM%*%CM; Sm<-S*0;  nonzero<-diag(CM%*%CM)==1;  LM<-S[nonzero,nonzero]; 
  if(sum(nonzero)==0) LM<-0*LM else LM<-solve(LM)  ## Inverse
  Sm[nonzero,nonzero]<-LM; return(Sm) ## Substitute inverse in result SM and return
}

## Evaluation of the criterion L at C,\lambda
Lhat<-function(XM,YM,CM,lambda, SM=pseudo(XM,CM), CXTY=CM%*%t(XM)%*%YM) 
     -sum(SM)/2*lambda^2+ sum(SM%*%CXTY)*lambda+sum(YM^2)/2-t(CXTY)%*%SM%*%CXTY/2 

## Compute all possible candidate shrinkage moves at orthant CM and \lambda = Lm
shrink<-function(XM,YM,CM,Lm,TOL=10){  ## Variables for results, S^-, S^-CXTY, S^- 1
  ucc<-lambdacc<-result<-c(); SM<-pseudo(X=XM,C=CM);  SMCXTY<-SM%*%CM%*%t(XM)%*%YM;  
  SM1<-apply(X=SM,MARGIN = 1,FUN=sum )
  lambdacc<-round(SMCXTY/SM1,TOL); ## << Equation (9) to compute candidate lambda >>
  for(lj in lambdacc) ucc<-cbind(ucc, round(SMCXTY-lj*SM1,TOL) )  ## The candidate \hat{u} solutions
  ### Filter results:  clear NA, Inf, \beta<0, <=\lambda  ;   apply filter and adapt size
  filtroc<-(!is.na(lambdacc)) &  (!is.infinite(lambdacc)) & (apply(ucc>=0,2,prod)==1) & (lambdacc>Lm);
  lambdacc<-lambdacc[filtroc];  ucc<-ucc[,filtroc];  if(length(ucc)==ncol(XM)) ucc<-matrix(ncol=1,ucc)  
  if(length(lambdacc)>=1) ## For valid lambda, compute \beta,L from every column \hat{u}
    for(ik in 1:ncol(ucc)) 
      result<-rbind(result, c(lambdacc[ik], CM%*%matrix(ncol=1,ucc[,ik]), 
                              Lhat(XM=XM,YM=YM,CM=CM,lambda=lambdacc[ik])  )   ) ## {lambda,beta,L}
  return(result)
}
  
## Lasso by orthants
lassoq<-function(XM,YM, TOL=10){  ## Initialization, beta_ols, matrix C 
  res<-c(); Lm<-0; p<-ncol(XM); Beta0<-lm(YM~XM-1); ## lambda, number of variables, Beta_ols
  Cm<-diag(sign(round(Beta0$coefficients,TOL))); Cm[is.na(diag(Cm)),is.na(diag(Cm))]<-0
  res<-c(Lm,Beta0$coefficients,sum(Beta0$residuals^2)/2)  ##  initial step (lambda=0, beta, L)
  while(!identical(diag(Cm),rep(0,p))){    ###################### main loop
    ## candidates to shrink, reactivate, temporary results
    jc<-(1:p)[diag(Cm%*%Cm)!=0]; jcc<-(1:p)[diag(Cm%*%Cm)==0]; cand<-c()
    if(length(jc)<p)  ### If there are zeroes in C, first try to reactivate
      for(kk in jcc) ## kk indexes which variable to reactivate
        for(candvalue in c(-1,1)){ ## candvalue gives -+1 signs, pdate C' to reactivate
          Cmc<-Cm; Cmc[kk,kk]<-candvalue; cand<-rbind(cand,shrink(XM = XM,YM = YM,CM = Cmc,Lm=Lm,TOL=TOL))
          }   ## end of -+1 loop, end of reactivate  
    ## then perform  shrinkage step
    Cm->Cmc;  cand<-rbind(cand,shrink(XM = XM,YM = YM,CM = Cmc,Lm=Lm,TOL=TOL))
    ## Using the reactivation/shrinkage results, select the next move
    Ind<-which.min(cand[,1]);  Lm<-cand[Ind,1] ## select smallest lambda, update Lm
    res<-rbind(res, cand[Ind,] ); Cm	<-diag(sign(cand[Ind,1+1:p])) ## update path, orthant
  }  #################### end of main loop
  return(unname(res)); ## output is (lambda, beta, L)
}
\end{verbatim}
}

As example of the \texttt{lassoq} code, we compute the path 
  for data of Table \ref{tab_ex1d}(a)
and reproduce Table \ref{tablepath}(a). The code needs preloading the 
functions of this Appendix.

\begin{knitrout}
\definecolor{shadecolor}{rgb}{0.969, 0.969, 0.969}\color{fgcolor}\begin{kframe}
\begin{alltt}
\hlstd{X}\hlkwb{<-}\hlkwd{c}\hlstd{(}\hlnum{0}\hlstd{,}\hlnum{0}\hlstd{,}\hlopt{-}\hlnum{1}\hlstd{,}\hlnum{1}\hlstd{,}\hlopt{-}\hlnum{1}\hlstd{,}\hlnum{1}\hlstd{,}\hlnum{0}\hlstd{,}\hlnum{1}\hlstd{,}\hlnum{0}\hlstd{,}\hlopt{-}\hlnum{1}\hlstd{,}\hlopt{-}\hlnum{1}\hlstd{,}\hlnum{0}\hlstd{,}\hlopt{-}\hlnum{1}\hlstd{,}\hlnum{0}\hlstd{,}\hlnum{0}\hlstd{,}\hlopt{-}\hlnum{1}\hlstd{,}\hlopt{-}\hlnum{1}\hlstd{,}\hlnum{1}\hlstd{,}\hlnum{0}\hlstd{,}\hlnum{1}\hlstd{,}\hlopt{-}\hlnum{1}\hlstd{,}\hlopt{-}\hlnum{1}\hlstd{,}\hlopt{-}\hlnum{1}\hlstd{,}\hlnum{1}\hlstd{,}\hlnum{4}\hlstd{,}\hlnum{0}\hlstd{,}\hlnum{3}\hlstd{,}\hlopt{-}\hlnum{3}\hlstd{)}
\hlstd{X}\hlkwb{<-}\hlkwd{matrix}\hlstd{(X,}\hlkwc{byrow}\hlstd{=}\hlnum{TRUE}\hlstd{,}\hlkwc{ncol}\hlstd{=}\hlnum{4}\hlstd{); XM}\hlkwb{<-}\hlstd{X[,}\hlopt{-}\hlnum{4}\hlstd{];   YM}\hlkwb{<-}\hlkwd{matrix}\hlstd{(}\hlkwc{ncol}\hlstd{=}\hlnum{1}\hlstd{,X[,}\hlnum{4}\hlstd{])}
\hlkwd{lassoq}\hlstd{(}\hlkwc{XM}\hlstd{=XM,}\hlkwc{YM}\hlstd{=YM)} \hlcom{## Columns are lambda, betas, L; each row a breakpoint}
\end{alltt}
\begin{verbatim}
##            [,1]       [,2]      [,3]       [,4]      [,5]
## [1,]  0.0000000  0.1142857 0.8714286 -1.1857143 0.8428571
## [2,]  0.1176471  0.0000000 0.7352941 -1.0294118 1.0743945
## [3,]  0.3333333  0.0000000 0.6666667 -1.0000000 1.4444444
## [4,]  1.4186047 -0.3720930 0.0000000 -0.3953488 2.7652785
## [5,]  5.4285714 -0.4285714 0.0000000  0.0000000 5.1632653
## [6,] 14.0000000  0.0000000 0.0000000  0.0000000 7.0000000
\end{verbatim}
\end{kframe}
\end{knitrout}

The function \texttt{lassoqw} is a simple adaptation of \texttt{lassoq} 
to perform adaptive lasso. We 
perform the analysis of the same data with weights $w_i=1/|\hat\beta_i^{OLS}|^\gamma$, where 
$\gamma$ is fixed and 
$\hat\beta_i^{OLS}$ is the $i$-th coefficient of the least squares 
fit to the data. We give results below for $\gamma=0.25,1$.

\begin{knitrout}
\definecolor{shadecolor}{rgb}{0.969, 0.969, 0.969}\color{fgcolor}\begin{kframe}
\begin{alltt}
\hlkwd{lassoqw}\hlstd{(}\hlkwc{XM}\hlstd{=XM,}\hlkwc{YM}\hlstd{=YM,}\hlkwc{adaptive} \hlstd{=} \hlnum{TRUE}\hlstd{,}\hlkwc{gamma}\hlstd{=}\hlnum{0.25}\hlstd{)}
\end{alltt}
\begin{verbatim}
##             [,1]       [,2]      [,3]       [,4]      [,5]
## [1,]  0.00000000  0.1142857 0.8714286 -1.1857143 0.8428571
## [2,]  0.09594963  0.0000000 0.7414637 -1.0325815 1.0352911
## [3,]  1.03873325  0.0000000 0.4342734 -0.9060934 2.4139669
## [4,]  2.07061914 -0.1135323 0.0000000 -0.6283158 3.0895394
## [5,]  3.05595699  0.0000000 0.0000000 -0.6726211 3.9085728
## [6,] 11.47856765  0.0000000 0.0000000  0.0000000 6.9904572
\end{verbatim}
\begin{alltt}
\hlkwd{lassoqw}\hlstd{(}\hlkwc{XM}\hlstd{=XM,}\hlkwc{YM}\hlstd{=YM,}\hlkwc{adaptive} \hlstd{=} \hlnum{TRUE}\hlstd{,}\hlkwc{gamma}\hlstd{=}\hlnum{1}\hlstd{)}
\end{alltt}
\begin{verbatim}
##             [,1]      [,2]      [,3]       [,4]      [,5]
## [1,]  0.00000000 0.1142857 0.8714286 -1.1857143 0.8428571
## [2,]  0.03374469 0.0000000 0.7608727 -1.0411072 0.9141630
## [3,]  2.19961666 0.0000000 0.0000000 -0.7620751 3.7633227
## [4,] 13.04285714 0.0000000 0.0000000  0.0000000 6.8261139
\end{verbatim}
\end{kframe}
\end{knitrout}

\subsection*{Appendix 5 - Elastic net \texttt{R} code and example}
            
The code has the same structure of orthant lasso:
Algorithm \ref{algo} is implemented in main function \texttt{elastiq}.
The shrinking step  \ref{SE1} of Algoritm \ref{algostep} 
is the call to the 
numerical solution of Equation (\ref{eq_enets}), with two 
alternatives given: bisection and secant 
implementations in \texttt{bisect} and \texttt{secant}. The
rest of functions are \texttt{pseudomu} and \texttt{SM} to
compute $\mathbf{S}(\lambda)^-$; \texttt{C2u} for $\mathbf{C}^2\hat{\mathbf{u}}$ 
of Equation (\ref{ec_c2unet}) and 
\texttt{Ehat} to compute $\hat E_\mathbf{C}$.

This code is provided without accepting any responsibility 
for its accuracy, use or misuse of code or results.

{\scriptsize
\begin{verbatim}
## Pseudo inverse of CX^TXC + \mu C^2,  here C is diagonal of {+-1,0} entries
pseudomu<-function(XM,CM,mu){
  S<-CM%*%t(XM)%*%XM%*%CM + mu*CM%*%CM; Sm<-S*0  ## big matrix
  nonzero<- diag(CM%*%CM)==1;  LM<-S[nonzero,nonzero] ## invertible submatrix
  if(sum(nonzero)==0) LM<-LM*0 else LM<-solve(LM)
  Sm[nonzero,nonzero]<-LM; return(Sm) ## Substitute inverse in result Sm and return
}

### Call to pseudomu() to compute generalized inverse S(lambda)^-
SM<-function(XM,CM,alpha=0.5,lambda)  pseudomu(XM=XM,CM=CM,mu=lambda*(1-alpha))
## Evaluate C^2\hat{u}
C2u<-function(XM,YM,CM,alpha=0.5,lambda)  
  SM(XM=XM,CM=CM,alpha=alpha,lambda = lambda)%*%(CM%*%t(XM)%*%YM - alpha*lambda)
## Evaluation of criterion \hat E_C, i.e. value of E at \hat{beta}=CC^2\hat{u}=C\hat{u}
Ehat<-function(CM,XM,YM,lambda,alpha,betahat=CM%*%C2u(XM=XM,YM=YM,CM=CM,alpha=alpha,lambda=lambda))
   sum((YM-XM%*%betahat)^2)/2+lambda*alpha*sum(t(betahat)%*%CM)+lambda*(1-alpha)*sum(betahat^2)/2  

## Secant to solve Equation (12) for \lambda
secant<-function(XM,YM,CM,alpha=0.5,lambda=0,l1=1.01*lambda+0.01,lhigh=max(abs(t(XM)%*%YM))/alpha,
                 Nmax=15,TOL=6,ii=1){
  i<-1; l0<-lambda; rango<-c(0, lhigh); FLAG<-TRUE; ## counter, lambda values and exit flag
  while(FLAG){
    lnew<-l1 - C2u(XM=XM,YM=YM,CM=CM,alpha=alpha,lambda = l1)[ii] * (l1-l0) / 
      ( C2u(XM=XM,YM=YM,CM=CM,alpha=alpha,lambda = l1)[ii]-C2u(XM=XM,YM=YM,CM=CM,alpha=alpha,lambda = l0)[ii]   )
    l0<-l1; l1<-lnew;  i<-i+1 ## update lambda values then exit conditions
    if(i>Nmax) FLAG<-FALSE
    if( abs(l0-l1)<10^-TOL   ) FLAG<-FALSE
    if(  (abs(lnew)>10*max(rango))|| (lnew<0)){
      FLAG<-FALSE;  l1<-100*abs(lnew)
    }
  }
  if( prod(round(C2u(XM=XM,YM=YM,CM=CM,alpha=alpha,lambda = l1),TOL)>=0 )==0) l1<--1 ## check C2u>0
  return(l1)
}

## Bisection to solve Equation (12) for \lambda
bisect<-function(XM,YM,CM,alpha=0.5,lambda=0,llow=lambda,lhigh=max(abs(t(XM)%*%YM))/alpha,  
                 Nmax=55,TOL=6,ii=1){ ## initial values
  i<-1; FLAG<-TRUE ## index, exit flag
  while(FLAG){ ## try and vtry are values of lambda and the respective ii-th coordinate of C2u 
    try<-c(llow,mean(c(llow,lhigh)),lhigh); vtry<-c(); 
    for(l in try) vtry<-c(vtry, C2u(XM=XM,YM=YM,CM=CM,alpha=alpha,lambda = l)[ii])
    vtry<-sign(vtry);   i<-i+1; ## update index, lambdas to try then exit conditions
    if( prod(vtry[1:2])==-1 ) lhigh<-try[2] else llow<-try[2]
    if( abs(llow-lhigh)<10^-TOL   ) FLAG<-FALSE
    if(i>Nmax) FLAG<-FALSE
  }  
  l1<-mean(try) ## Candidate lambda
  if( prod(round(C2u(XM=XM,YM=YM,CM=CM,alpha=alpha,lambda = l1),TOL)>=0 )==0) l1<--1 ## check C2u>0
  return(l1)
}

## Elastic net by orthants
elastiq<-function(XM,YM,TOL=8,alpha=0.5){## Initialization
Beta0<-lm(YM~XM-1); Cm<-diag(sign(Beta0$coefficients)); p<-ncol(XM)
res<-matrix(nrow=1, round( c(0,Beta0$coefficients,Ehat(CM=Cm,XM=XM,YM=YM,lambda=0,alpha=alpha)) , TOL) )
while(!identical( diag(Cm),rep(x=0,times=p))){   ###### main loop
  cand<-c()
  for(i in 1:p){
    if(Cm[i,i]==0){ ## first try to reactivate 
      for(k in c(-1,1)){
        Cmc<-Cm;  Cmc[i,i]<-k
        for(j in (1:p)[diag(Cmc)!=0]){  ## shrink all coordinates in new orthant
        ##secant(XM=XM,YM=YM,CM=Cmc,alpha=alpha,ii=j,TOL=TOL)->lambdac ## uncomment one method
        bisect(XM=XM,YM=YM,CM=Cmc,alpha=alpha,ii = j,llow = max(res[,1]),TOL=TOL)->lambdac ## uncomment one method
        Cmc%*%C2u(XM=XM,YM=YM,CM=Cmc,alpha=alpha,lambda = lambdac)->betatemp
        cand<-rbind(cand,c(lambdac,betatemp,Ehat(CM=Cmc,XM=XM,YM=YM,lambda=lambdac,alpha=alpha)))  ## 1 reactivate
        }
      }
    } else  { ## then perform shrinkage moves
      Cmc<-Cm;  #Cmf[i,i]<-0; #Cmc[i,i]<-0;
      ##secant(XM=XM,YM=YM,CM=Cmc,alpha=alpha,ii=i,TOL=TOL)->lambdac ## uncomment one method
      bisect(XM=XM,YM=YM,CM=Cmc,alpha=alpha,ii = i,llow = max(res[,1]),TOL=TOL)->lambdac ## uncomment one method
      Cmc%*%C2u(XM=XM,YM=YM,CM=Cmc,alpha=alpha,lambda = lambdac)->betatemp
      cand<-rbind(cand,c(lambdac,betatemp,Ehat(CM=Cmc,XM=XM,YM=YM,lambda=lambdac,alpha=alpha))) ## -1 shrink
    }
  }
  cand<-round(cand,TOL);  if(!is.matrix(cand)) cand<-matrix(nrow=1,cand)
  ## Remove $\lambda$ that repeat past steps then select smallest valid \lambda, update results and orthant
  ff<-cand[,1]>max(res[,1])+10*10^-(TOL); cand<-cand[ff,];   if(!is.matrix(cand)) cand<-matrix(nrow=1,cand) 
  Ind<-which.min(cand[,1]); res<-rbind(res, cand[Ind,]);   diag(Cm)<-sign(cand[Ind,1+1:p])
} ###### end of main loop
return(unname(res))
}
\end{verbatim}
}

\pagebreak

The example uses \texttt{elastiq} with $\alpha=0.5$ for the 
data of Table \ref{tab_ex1d}(a).
We use the provided \texttt{R} functions, and \texttt{XM} and \texttt{YM} of Appendix 4
 to reproduce Table \ref{tablepath}(b). We also give another case of elastic net for the
 same data and $\alpha=0.9$.

\begin{knitrout}
\definecolor{shadecolor}{rgb}{0.969, 0.969, 0.969}\color{fgcolor}\begin{kframe}
\begin{alltt}
\hlcom{## Columns are lambda, betas, L; each row a breakpoint}
\hlkwd{elastiq}\hlstd{(}\hlkwc{XM}\hlstd{=XM,}\hlkwc{YM}\hlstd{=YM,}\hlkwc{TOL}\hlstd{=}\hlnum{8}\hlstd{)}
\end{alltt}
\begin{verbatim}
##            [,1]       [,2]      [,3]       [,4]      [,5]
## [1,]  0.0000000  0.1142857 0.8714286 -1.1857143 0.8428571
## [2,]  0.1459742  0.0000000 0.7315377 -1.0262653 1.0539203
## [3,]  0.2471659  0.0000000 0.7039861 -1.0132639 1.1811668
## [4,]  2.6872073 -0.3743399 0.0000000 -0.3589718 2.8979158
## [5,] 16.9614814 -0.1937892 0.0000000  0.0000000 6.4652136
## [6,] 28.0000000  0.0000000 0.0000000  0.0000000 7.0000000
\end{verbatim}
\begin{alltt}
\hlkwd{elastiq}\hlstd{(}\hlkwc{XM}\hlstd{=XM,}\hlkwc{YM}\hlstd{=YM,}\hlkwc{TOL}\hlstd{=}\hlnum{8}\hlstd{,}\hlkwc{alpha}\hlstd{=}\hlnum{0.9}\hlstd{)}
\end{alltt}
\begin{verbatim}
##            [,1]       [,2]      [,3]       [,4]      [,5]
## [1,]  0.0000000  0.1142857 0.8714286 -1.1857143 0.8428571
## [2,]  0.1223731  0.0000000 0.7346599 -1.0288828 1.0709295
## [3,]  0.3125817  0.0000000 0.6760267 -1.0032808 1.3801569
## [4,]  1.5631239 -0.3732292 0.0000000 -0.3900203 2.7791579
## [5,]  6.5623470 -0.3918375 0.0000000  0.0000000 5.4142556
## [6,] 15.5555555  0.0000000 0.0000000  0.0000000 7.0000000
\end{verbatim}
\end{kframe}
\end{knitrout}

\pagebreak
\subsection*{Appendix 6 - Additional figures}

\begin{figure}[!h]
\begin{center}
\begin{knitrout}
\definecolor{shadecolor}{rgb}{0.969, 0.969, 0.969}\color{fgcolor}
\includegraphics[width=\maxwidth]{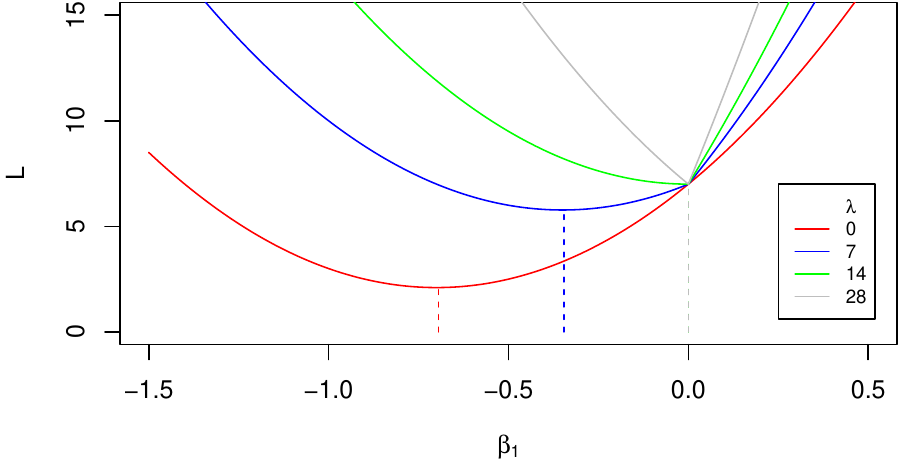} 
\end{knitrout}
\end{center}
\caption{Lasso criterion $L$ for Example \ref{ex_1d3}.} \label{figex1d}
\end{figure}

\begin{figure}
\begin{tabular}{cc}
\begin{knitrout}
\definecolor{shadecolor}{rgb}{0.969, 0.969, 0.969}\color{fgcolor}
\includegraphics[width=\maxwidth]{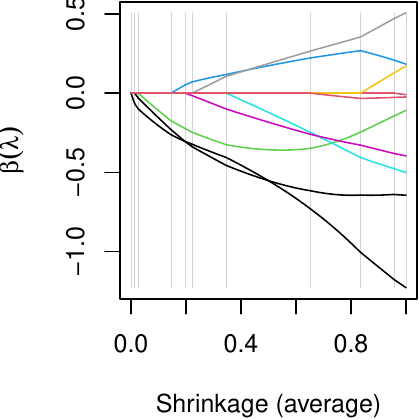} 
\end{knitrout}
&
\begin{knitrout}
\definecolor{shadecolor}{rgb}{0.969, 0.969, 0.969}\color{fgcolor}
\includegraphics[width=\maxwidth]{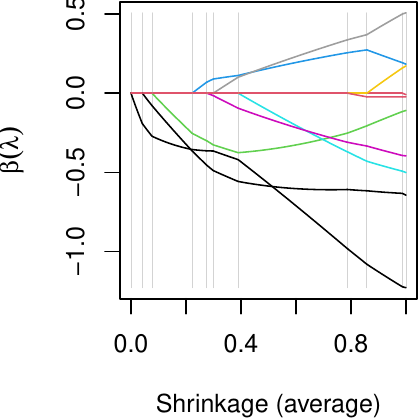} 
\end{knitrout}
\\
(a)&(b)\end{tabular}
\caption{(a) and (b) elastic net paths for synthetic data and $\alpha$ values $0.4$ and $0.8$,
respectively. The horizontal shrinkage in the plots is the average of $L_1$ and $L_2$ norms. }\label{fig_lasso_netex2}
\end{figure}

\begin{figure}[!h]
\begin{center}
\begin{tabular}{cc}
\begin{minipage}[t]{2.8in}
\begin{scriptsize} 
\begin{spacing}{0.6}
\texttt{-+-+++-++--++------++---+----+--+-++---+----+\\ -+-+++-++--++------0+---+----+--+-++---+----+\\ -+-+++-++--++-------+---+----+--+-++---+----+\\ -+-+++-++-0++-------+---+----+--+-++---+----+\\ -+-+++-++-0++-------+---+----+--+-0+---+----+\\ -+-+++-++-0++-------+---+----+--+--+---+----+\\ -+-+++-++-0++-------+---+--0-+--+--+---+----+\\ -+-+++-++-0++-------+---+-00-+--+--+---+----+\\ -+-+++-+0-0++-------+---+-00-+--+--+---+----+\\ -+-+++-+0-0++-------+---+-00-+--+--+--0+----+\\ -+-+++-+0-0++-------+---+-00-+-0+--+--0+----+\\ -+-+++-+0-0++-------+---+-0+-+-0+--+--0+----+\\ -+-+++-+0-0++-------+---+-0+-+00+--+--0+----+\\ -+-+++-+0-0++-0-----+---+-0+-+00+--+--0+----+\\ -+-+++-+0-0++-0-----+---+-0+-+00+--+--0+---0+\\ -+-+++-+0-0++-0-----+---+-0+-+00+--+--++---0+\\ -+-+++-+0-0++00-----+---+-0+-+00+--+--++---0+\\ -+-+++-+0-0++00-----+---+-0+-+00+--0--++---0+\\ -+-+++-+0-0++00-----+---+-0+-+00+--0--+0---0+\\ -+-+++-+0-0++00-----+---+-0+0+00+--0--+0---0+\\ -+-+++-+0-0++00-----+---0-0+0+00+--0--+0---0+\\ -+-+++-+0-00+00-----+---0-0+0+00+--0--+0---0+\\ -+-+++-+0-00+00-----+---0-0+0+00+--0--00---0+\\ -+-+++-+0-00+00-----+---0-0+0+00+--00-00---0+\\ -+-+++-+0-00+00-----+-0-0-0+0+00+--00-00---0+\\ -+-+++-+0-00+00-----+-0-000+0+00+--00-00---0+\\ -+-+++-+0-00+00-----+-0-000+0+00+--00-00--00+\\ -+-+++-+0-00+00-----+-0--00+0+00+--00-00--00+\\ -+-+++-+0-00+00-----+-0--00+0+00+-000-00--00+\\ -+-+++-+0-00+00--0--+-0--00+0+00+-000-00--00+\\ -+-+++0+0-00+00--0--+-0--00+0+00+-000-00--00+\\ -+-+0+0+0-00+00--0--+-0--00+0+00+-000-00--00+\\ -+-+0+0+0-00+00--0--+-0--0000+00+-000-00--00+\\ -+-+0+0+0-00+00--0--+-0--0000+00+0000-00--00+\\ -+-+0+0+0-00+00--0--+-0--0000+00+0000-00--000\\ -+-+0+0+0-00000--0--+-0--0000+00+0000-00--000\\ -+-+0+0+0-000000-0--+-0--0000+00+0000-00--000\\ -+-00+0+0-000000-0--+-0--0000+00+0000-00--000\\ -+-00+0+0-000000-0--+-00-0000+00+0000-00--000\\ -+-00+0+0-000000-0--+-00-0000+00+0000-000-000\\ -+-00+0+0-000000-0--+-00-0000+00+00000000-000\\ -+000+0+0-000000-0--+-00-0000+00+00000000-000\\ -+000+0+0-000000-0--+-0000000+00+00000000-000\\ -+000+0+0-000000-0--+-0000000000+00000000-000\\ -+000+0+0-000000-0--0-0000000000+00000000-000\\ -+000+0+0-000000-0--0-0000000000+000000000000\\ -+000+0+0-000000-00-0-0000000000+000000000000\\ -+000+0+0-000000000-0-0000000000+000000000000\\ -+00000+0-000000000-0-0000000000+000000000000\\ -+00000+00000000000-0-0000000000+000000000000\\ -+00000000000000000-0-0000000000+000000000000\\ -+00000000000000000-0-00000000000000000000000\\ -+00000000000000000-0000000000000000000000000\\ -+0000000000000000000000000000000000000000000\\ -00000000000000000000000000000000000000000000\\ 000000000000000000000000000000000000000000000}
\end{spacing}
\end{scriptsize}
\end{minipage}&
\begin{minipage}[t]{2.8in}
\begin{scriptsize} 
\begin{spacing}{0.6}
\texttt{-+-+++-++--++------++---+----+--+-++---+----+\\ -+-+++-++--++------0+---+----+--+-++---+----+\\ -+-+++-++--++-------+---+----+--+-++---+----+\\ -+-+++-++-0++-------+---+----+--+-++---+----+\\ -+-+++-++-0++-------+---+----+--+-0+---+----+\\ -+-+++-++-0++-------+---+----+--+--+---+----+\\ -+-+++-++-0++-------+---+--0-+--+--+---+----+\\ -+-+++-++-0++-------+---+-00-+--+--+---+----+\\ -+-+++-+0-0++-------+---+-00-+--+--+--0+----+\\ -+-+++-+0-0++-------+---+-00-+-0+--+--0+----+\\ -+-+++-+0-0++-------+---+-0+-+-0+--+--0+----+\\ -+-+++-+0-0++-0-----+---+-0+-+00+--+--0+----+\\ -+-+++-+0-0++-0-----+---+-0+-+00+--+--++----+\\ -+-+++-+0-0++-0-----+---+-0+-+00+--+--++---0+\\ -+-+++-+0-0++00-----+---+-0+-+00+--+--++---0+\\ -+-+++-+0-0++00-----+---+-0+-+00+--0--+0---0+\\ -+-+++-+0-0++00-----+---+-0+0+00+--0--+0---0+\\ -+-+++-+0-0++00-----+---0-0+0+00+--0--+0---0+\\ -+-+++-+0-00+00-----+---0-0+0+00+--0--00---0+\\ -+-+++-+0-00+00-----+---0-0+0+00+--00-00---0+\\ -+-+++-+0-00+00-----+-0-000+0+00+--00-00---0+\\ -+-+++-+0-00+00-----+-0-000+0+00+--00-00--00+\\ -+-+++-+0-00+00-----+-0-000+0+00+-000-00--00+\\ -+-+++-+0-00+00--0--+-0--00+0+00+-000-00--00+\\ -+-+0+-+0-00+00--0--+-0--00+0+00+0000-00--00+\\ -+-+0+0+0-00+00--0--+-0--0000+00+0000-00--000\\ -+-00+0+0-00000--0--+-0-00000+00+0000-00--000\\ -+-00+0+0-00000--0--+-0000000+00+0000-000-000\\ -+000+0+0-000000-0--+-0000000+00+0000-000-000\\ -+000+0+0-000000-0--+-0000000+00+00000000-000\\ -+000+0+0-000000-0--0-0000000000+000000000000\\ -+000+0+0-00000000--0-0000000000+000000000000\\ -+000+0+0-000000000-0-0000000000+000000000000\\ -+00000+0-000000000-0-0000000000+000000000000\\ -+0000000-000000000-0-0000000000+000000000000\\ -+00000000000000000-0-00000000000000000000000\\ -+0000000000000000000-00000000000000000000000\\ -+0000000000000000000000000000000000000000000\\ -00000000000000000000000000000000000000000000\\ 000000000000000000000000000000000000000000000}
\end{spacing}
\end{scriptsize}
\end{minipage}\\(a)&(b)
\end{tabular}
\end{center}
\caption{Orthant excursion for simulated data with $n=55$ observations and
$p=45$ variables, computed with (a) the orthant method and (b) \texttt{glmnet}.}\label{fig_quads}
\end{figure}

\end{document}